\documentclass[10pt,journal,compsoc]{IEEEtran}

\IEEEoverridecommandlockouts
\usepackage{url}
\usepackage{enumitem}
\usepackage{array}
\usepackage{algorithm}
\usepackage{algorithmic}
\usepackage{amsmath,amssymb,amsfonts}
\usepackage{balance}
\usepackage{bm}
\usepackage{cite}
\usepackage{graphicx}
\usepackage{subfig}
\usepackage{booktabs}
\usepackage[flushleft]{threeparttable}
\usepackage{multirow}
\usepackage{listings}
\usepackage{color}
\usepackage{framed}

\usepackage{footnote}

\newcommand{\rqbox}[1]{\begin{tcolorbox}[left=4pt,right=4pt,top=4pt,bottom=4pt,colback=gray!5,colframe=gray!40!black,before skip=6pt,after skip=6pt]#1\end{tcolorbox}}

\definecolor{jkcolor}{RGB}{0, 181, 29}

\usepackage{soul}
\usepackage{marvosym}
\usepackage{textcomp}
\usepackage{tcolorbox} 

\definecolor{mycolor}{RGB}{10, 186, 181}

\graphicspath{ {figures/} }

\definecolor{javared}{rgb}{0.6,0,0} 
\definecolor{javagreen}{rgb}{0.25,0.5,0.35} 
\definecolor{javapurple}{rgb}{0.5,0,0.35} 
\definecolor{javadocblue}{rgb}{0.25,0.35,0.75} 
\definecolor{lineno}{rgb}{0.5,0.5,0.5} 

\begin{document}

\title{Broken External Links on Stack Overflow}

 \author{Jiakun~Liu,
        Xin~Xia,
        David~Lo,
        Haoxiang~Zhang,
        Ying Zou,
        Ahmed~E.~Hassan,
        and Shanping~Li
\IEEEcompsocitemizethanks{\IEEEcompsocthanksitem Jiakun Liu and Shanping Li are with the College of Computer Science and Technology, Zhejiang University, Hangzhou, China. \protect\\
E-mail: \{jkliu, shan\}@zju.edu.cn
\IEEEcompsocthanksitem Xin Xia is with the Faculty of Information Technology, Monash University, Melbourne, Australia. \protect\\
E-mail: xin.xia@monash.edu
\IEEEcompsocthanksitem Haoxiang Zhang and Ahmed E. Hassan are with the Software Analysis and
Intelligence Lab (SAIL), Queen's University, Kingston, Ontario, Canada. \protect\\
E-mail: \{hzhang, ahmed\}@cs.queensu.ca
\IEEEcompsocthanksitem Ying Zou is with the Department of Electrical and Computer Engineering, Queen's University, Kingston, Ontario, Canada. \protect\\
E-mail: ying.zou@queensu.ca
\IEEEcompsocthanksitem David Lo is with the School of Information Systems, Singapore Management University, Singapore. \protect\\
E-mail: davidlo@smu.edu.sg
\IEEEcompsocthanksitem Xin Xia is the corresponding author.jkliu@zju.edu.cn 
}}

\pagenumbering{arabic}

\IEEEtitleabstractindextext{

\begin{abstract}
    Stack Overflow hosts valuable programming-related knowledge with 11,926,354 links that reference to the third-party websites.
    The links that reference to the resources hosted outside the Stack Overflow websites extend the Stack Overflow knowledge base substantially.
    However, with the rapid development of programming-related knowledge, many resources hosted on the Internet are not available anymore.
    Based on our analysis of the Stack Overflow data that was released on Jun. 2, 2019, 14.2\% of the links on Stack Overflow are broken links.
    The broken links on Stack Overflow can obstruct viewers from obtaining desired programming-related knowledge, and potentially damage the reputation of the Stack Overflow as viewers might regard the posts with broken links as obsolete.
    In this paper, we characterize the broken links on Stack Overflow.
    65\% of the broken links in our sampled questions are used to show examples, e.g., code examples.
    70\% of the broken links in our sampled answers are used to provide supporting information, e.g., explaining a certain concept and describing a step to solve a problem.
    Only 1.67\% of the posts with broken links are highlighted as such by viewers in the posts' comments.
    Only 5.8\% of the posts with broken links removed the broken links.
    Viewers cannot fully rely on the vote scores to detect broken links, as broken links are common across posts with different vote scores.
    The websites that host resources that can be maintained by their users are referenced by broken links the most on Stack Overflow -- a prominent example of such websites is GitHub.
    The posts and comments related to the web technologies, i.e., JavaScript, HTML, CSS, and jQuery, are associated with more broken links.
    Based on our findings, we shed lights for future directions and provide recommendations for practitioners and researchers.
    
    \end{abstract}


}

\IEEEdisplaynontitleabstractindextext
\IEEEpeerreviewmaketitle

\maketitle

\section{Introduction}

Stack Overflow is a valuable knowledge base that serves millions of users around the world \cite{rahman2014towards, xia2017developers}.
When developers communicate on Stack Overflow, they can use links to introduce the resources that are scattered across the Internet \cite{referencing, formatting}.
Based on the Stack Overflow data dump (released on Jun. 2, 2019), among 19,200,512 posts (i.e., questions and answers) and comments, 11,926,354 distinct links are referenced 27,553,546 times in total.

\begin{figure}[!htb]
    \centering
    \includegraphics[width = \linewidth]{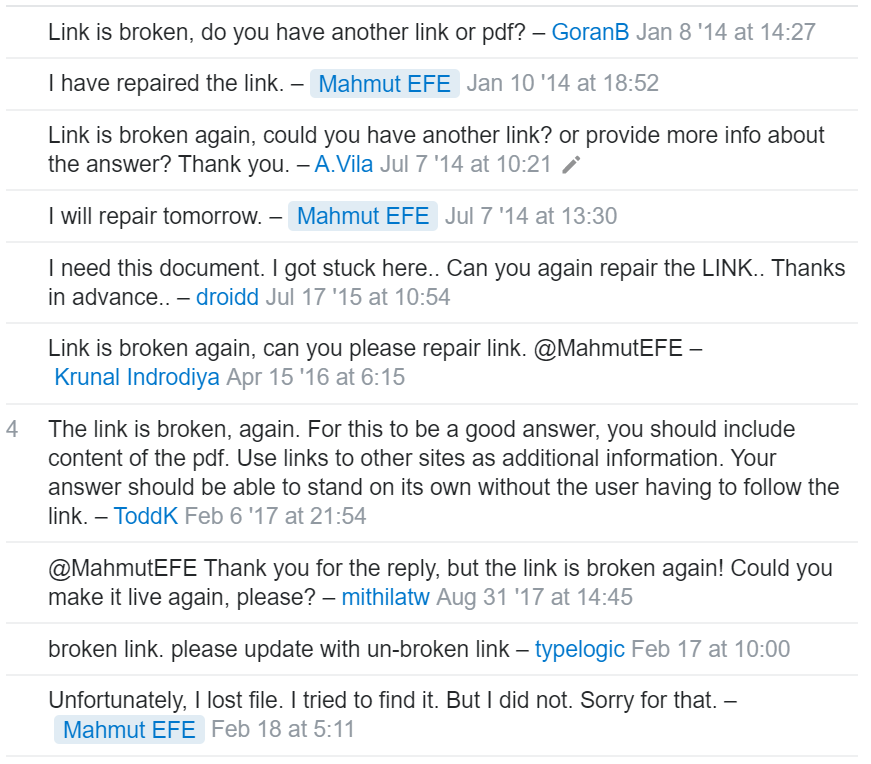}
    \caption{An example of the comments to the accepted answer with a broken link.
    The accepted answers can provide values to questioners to solve their problems.
    However, because viewers cannot fully understand the answer with broken links, the accepted answers with broken links cannot provide values to the future viewers.}\label{figure_example}
\end{figure}

However, with the rapid development of programming-related knowledge, many resources hosted on the Internet are not available anymore.
In this paper, we refer to the links that reference to unavailable resources as \textbf{broken links}.
In a prior study, Zhang et al. focused on analyzing obsolete knowledge on Stack Overflow and they observed that 11\% links in answers are broken links \cite{Zhang_2019}.
However, they did not analyze all the broken links on Stack Overflow.
Considering a large number of external resources that are referenced by Stack Overflow, it is unclear how Stack Overflow suffers from the broken link problems.
Figure \ref{figure_example} shows an example of the comments to the accepted answer that solves the authentication error for a Mifare card\footnote{https://stackoverflow.com/q/15881962/}.
Six comments that were received from Jan 08, 2014, to Feb 17, 2020, complain about the broken link.
Broken links can obstruct viewers from getting desired programming-related crowdsourced knowledge on Stack Overflow, and potentially damage the reputation of the Stack Overflow as viewers might regard the posts with broken links as obsolete \cite{Zhang_2019}.
Therefore, it is important to investigate the broken links on Stack Overflow and understand their impacts and characteristics.
By doing so, we could provide insights for practitioners and researchers to address this issue.

In our paper, we investigate the broken links on Stack Overflow.
To do so, we test the HTTP response status code (i.e., response code) of the 12,446,901 links in all versions of Stack Overflow posts.
To mitigate the intermittent behavior, we perform one test using a server located at Virginia, U.S. in Dec 2019, and another test using a server located at Singapore in Jan 2020.
We identify the links that are not responded with 2xx (e.g., 200, 201, 202) response code in both trials as broken links.
To understand the importance of investigating the broken links on Stack Overflow, we perform a series of preliminary studies on the broken links.
We observe that 14.2\% of the links are broken links.
404 response code is the most common response code for broken links on Stack Overflow.
Links that were posted earlier are more likely to be broken.
22.9\% of the links that were introduced in Aug 2008 (the month when the Stack Overflow website was established) are broken.
We structure our study by answering the following research questions:

\begin{enumerate}[wide=0pt]
    \item \textbf{What are the intended roles of the broken links on Stack Overflow?}\\
    65\% of the broken links in our sampled questions are used to show examples, e.g., code examples.
    70\% of the broken links in our sampled answers are used to provide supporting information, e.g., explaining a certain concept and describing a step to solve a problem.
    \item \textbf{What are the impacts of the broken links on Stack Overflow?}\\
    Only 1.57\% of the posts with broken links are highlighted as such by viewers in the posts' comments.
    Only 5.8\% of the posts with broken links repaired the broken links.
    Viewers cannot fully rely on the vote scores to detect broken links, as broken links are common across posts with different vote scores.
    \item \textbf{Which websites are referenced by broken links the most on Stack Overflow?}\\
    50\% of the broken links reference to the top 0.3\% websites ordered by the number of the broken links referencing to them.
    The websites that host the resources maintained by their users are referenced by broken links the most, e.g., github.com.
    \item \textbf{Are the posts and comments associated with particular tags more likely to have broken links than others?}\\
    55.4\% of the broken links are referenced in the posts and comments that are marked with the top 10 tags ordered by the number of the broken links associated with them.
    The posts and comments related to the web technologies, i.e., JavaScript, HTML, CSS, and jQuery, are associated with more broken links.
    
\end{enumerate}

Based on our findings, we provide actionable suggestions for Stack Overflow moderators, users, and future researchers.
For example, we encourage Stack Overflow to detect and mark the questions with broken links to notify viewers of the broken links.
For Stack Overflow users, we recommend Stack Overflow users to post the code in the code blocks or Stack Snippets as much as possible, rather than the external code websites, e.g., github.com.
For future researchers, we suggest they could repair broken links based on the revisions of links.

\noindent\textbf{Paper Organization:} The remainder of the paper is organized as follows.
Section \ref{s_rw} provides the background information of Stack Overflow and describes the related work about the knowledge sharing in software engineering and the broken links across the Internet.
Section \ref{setup} details our approach to collect and process the data which are used in our study.
Section \ref{s_pre} presents preliminary studies on the broken links on Stack Overflow.
Section \ref{findings} presents our research findings by answering the aforementioned four research questions.
Section \ref{dis} provides actionable suggestions based on our findings and acknowledges some of the key threats to the validity of our study.
Finally, Section \ref{con} concludes our study and proposes potential future work.

\section{Background and Related Work}\label{s_rw}

In this section, we present the background information of Stack Overflow and discuss the related work about the knowledge sharing in software engineering and broken links across the Internet.

\subsection{Studies on Stack Overflow}\label{ss_rw_mining_so}

Stack Overflow is a well-known online Q\&A site to answer programming-related problems.
Questioners can post questions that include textual descriptions \cite{howtoask}.
Each question may receive multiple answers \cite{howtoanswer}. 
Answers contribute the solutions to the crowdsourced knowledge on Stack Overflow.
Besides, registered viewers can comment under each post to notify the owner of the post for a clarification.
Many researchers focused on characterizing the Stack Overflow questions, answers and comments \cite{Saha:2013:TUC:2491411.2494585, Linares-Vasquez:2014:ACT:2597008.2597155}.
Saha et al. investigated why questions remain unanswered and concluded that the majority of them were due to low interest in the community \cite{Saha:2013:TUC:2491411.2494585}.
Linares-V'asquez et al. investigated the relationship between API changes in Android SDK and developers' reactions to those changes on Stack Overflow \cite{Linares-Vasquez:2014:ACT:2597008.2597155}.
They observed that Android developers usually have more questions when the behavior of APIs is modified.
Zhang et al. investigated the obsolete knowledge on Stack Overflow and observed that more than half of the obsolete answers were probably already obsolete when they were posted \cite{Zhang_2019}.
Zhang et al. observed that comments on Stack Overflow can be leveraged to improve the quality of their associated answers \cite{zhang2019comments}.

To capture the topics with which a question is associated, questioners need to specify the tags into well-defined categories when they post a question \cite{tagging}.
Each question can have at most five tags and must have at least one tag.
The tags can facilitate dispatching questions to the potential users who are interested in it.
Many researchers focused on the topics of the Stack Overflow questions \cite{Barua2014, Rosen2016, Bajaj:2014:MQA:2597073.2597083, wang2014entagrec, Xu_Linkable}.
They investigated the topic trends across the whole Stack Overflow \cite{Barua2014}, or in specific communities, e.g., mobile \cite{Rosen2016} and web \cite{Bajaj:2014:MQA:2597073.2597083}.
Xia et al. proposed an approach to recommend tags to software information sites \cite{xia2013tag}.
Xu et al. designed a tool to recognize semantically relevant knowledge units on Stack Overflow \cite{Xu_Linkable}.

Users can include code snippets and other references (e.g., links or images) to enrich their posted questions \cite{editing}.
Many researchers investigated how to utilizing the knowledge hosted on Stack Overflow to help with software engineering as well \cite{chen2015crowd, Cai_biker, Huang_biker}.
Chen et al. used the code blocks from Stack Overflow to detect defective code fragments in developers' source code \cite{chen2015crowd}.
Cai et al. and Huang et al. used the knowledge hosted on Stack Overflow to recommend APIs \cite{Cai_biker, Huang_biker}.

To ensure the quality of the crowdsourced knowledge on Stack Overflow, Stack Overflow propose a gamification system.
Questioners can accept the answers that can solve their questions \cite{accepted_answer}, i.e., can provide instant values to the developers who proposed questions on Stack Overflow.
Registered viewers could benefit from the accepted answers to learn the best way to solve the problems.
Registered viewers also can vote up the questions and answers that are useful to them, i.e., can also provide long-lasting values to the developers who encounter similar problems that are already asked on Stack Overflow \cite{vote_up, why_vote}.
By posting high-quality questions and answers, and suggesting reasonable edits, users can earn points to increase their reputations.
Considering the success of Stack Overflow, many researchers investigated the benefits of the gamification mechanisms \cite{Mamykina:2011:DLF:1978942.1979366, Cavusoglu_badge, Anderson:2012:DVC:2339530.2339665, pal2012evolution, Hanrahan_modeling_difficulties, li2015is}.
Anderson et al. designed a tool to determine which questions and answers are likely to have long-lasting value, and which ones are in need of additional help from the community \cite{Anderson:2012:DVC:2339530.2339665}.
Pal et al. investigated the evolution of experts on the Stack Overflow community and pointed out how expert users differ from ordinary users in terms of their contributions \cite{pal2012evolution}.
Hanrahan et al. developed indicators for difficult problems and experts \cite{Hanrahan_modeling_difficulties}.
They examined how complex problems are handled and dispatched across multiple experts.

Any user can suggest edits to revise a question's title, body, and tags, or an answer's body \cite{editing}.
Suggested edits from the original questioners and answerers will be applied immediately, as well as from users who have more than 2,000 reputation points (2k users).
Other users' suggested edits will be reviewed by the 2k users to decide whether to be applied or not.
Many researchers investigated the collaborative editing on Stack Overflow \cite{li2015is, wang2018users, chen_by_community, chen_data_driven}.
Li et al. observed that the benefits of collaborative editing on questions and answers outweigh its risks \cite{li2015is}.
Wang et al. observed that 25\% of the users did not make any more revisions once they received their first revision-related badge \cite{wang2018users}.
Chen et al. developed an edit-assistance tool to identify minor textual issues in posts and recommending sentence edits for correction \cite{chen_by_community}.
They also developed a Convolutional Neural Network-based approach to learn editing patterns from historical post edits for identifying the need for editing a post \cite{chen_data_driven}.
To characterize what do users (e.g., questioners, answers, and commenters) do after the observations of broken links, in this paper, we analyze the applied edits in this paper.
  
\subsection{Link Sharing in Software Engineering}\label{ss_rw_knowledge_sharing}

Researchers investigated the links in Stack Overflow.
Ye et al. investigated the \textbf{internal links} (i.e., links that reference to the resources hosted within the Stack Overflow website) to analyze the evolution of the knowledge network that is connected by the internal links \cite{Ye2017}.
G\'{o}mez et al. \cite{gomez2013study} investigated the \textbf{external links} (i.e., links that reference to the resources hosted outside the Stack Overflow website) from the link types, website types, and the most referenced links and websites perspectives.
Baltes et al. analyzed the purpose of the links that reference to documentation websites on Stack Overflow, e.g., pointing
to API documentation and concept descriptions on Wikipedia for background readings \cite{baltes2020contextual}.
Correa et al. investigated the role and impact of Stack Overflow in issue tracking systems \cite{correa2013integrating}.
They observed that the average number of comments posted in response to bug reports is less when Stack Overflow links are presented in the bug report.
Wang et al. revealed the links between the Android issues in bug tracking systems and Stack Overflow posts by integrating the semantic similarity between Android issues and Stack Overflow posts \cite{wang2015automatic}.

Researchers also investigated the links between software engineering artifacts.
For example, Rath et al. investigated the inter-linking of commits and issues in open source projects and observed that among six large projects, 60\% of the commits are linked to issues\cite{rath2018traceability}.

On utilizing web resources, Xia et al. listed the frequency and difficulty of the different web search tasks performed by developers \cite{xia2017developers}.
Rahman et al. proposed a novel IDE-based web search that exploits three reliable web search engines (e.g., Google, Bing, and Yahoo) and a programming Q\&A site (i.e., Stack Overflow) through their API endpoints \cite{rahman2014towards}.
Gao et al. developed an automatic web resources linking technique to linkify entity mentions to relevant official documentation in Stack Overflow \cite{8292806}.

Similar to the aforementioned studies, our work investigates knowledge dissemination in software engineering.
We focus on the broken links on Stack Overflow, rather than all the links, to study the broken link-sharing activities.

\subsection{Broken Links Across the Internet}

The HTTP response status code (i.e., response code) represents the result of the response of the website serve to the request of the link.
The response code is a three-digit integer.
The first digit of the status-code defines the class of response.
For example, 2xx response code represents the action requested by the client is received, understood, and accepted; 4xx response code represents that the request contains bad syntax or cannot be fulfilled; 5xx response code represents that the server failed to fulfill an apparently valid request \cite{fielding2014hypertext}.

Habibzadeh et al. examined the prevalence of the broken links in academic literature \cite{habibzadeh2013decay}.
They found that ranging 35.4\% to 68.4\% of the links in different journals are broken links.
Fetterly et al. observed that about one link out of every 200 broke each week on the Web\cite{fetterly2004large}.
Koehler et al. observed that the links could have dramatically different half-lives \cite{koehler2004longitudinal},
the links selected for publication appear to have greater longevity than the average links.
A 2015 study by Weblock analyzed more than 180,000 links from references in the full-text corpora of three major open-access publishers.
This study found that 24.5\% of the studied links are broken\footnote{https://web.archive.org/web/20160304081204/https://webloc-k.io/report?id=all}.
McCown et al. observed that half of the links cited in D-Lib Magazine articles were active 10 years after publication \cite{mccown2005availability}.
Hennessey et al. analyzed nearly 15,000 links in abstracts from Thomson Reuters's Web of Science citation index \cite{hennessey2013cross}.
They observed that the median lifespan of web pages was 9.3 years, and just 62\% were archived.
Klein et al. observed that one out of five Science, Technology, and Medicine articles suffering from reference rot, meaning it is impossible to revisit the web context that surrounds them after their publication\cite{klein2014scholarly}.
Zeng et al. observed that most resources linked in biomedical articles disappear in 8 years \cite{zeng2019dead}.

Different from the aforementioned studies, our study inspects the broken links in a popular software engineering related Q\&A websites, i.e., Stack Overflow.
Stack Overflow host a large collection of knowledge for developers to solve their programming-related problems.
Inspecting broken links on Stack Overflow could enable us to understand the broken links problems in the software engineering field.

\section{Experiment Setup}\label{setup}
In this section, we present the data collection steps that we used to extract the links from the SOTorrent dataset and test the availability of links.

\subsection{Data Collection}\label{ss_su_link_extraction}

Links on Stack Overflow can reference to the resources that are scattered across the Internet.
The links with \textit{stackoverflow.com} root domain reference to the resources hosted within the Stack Overflow websites.
In contrast, the links without \textit{stackoverflow.com} root domain reference to the resources hosted outside the Stack Overflow websites.
In this paper, we investigate the availability of the links that reference to the resources hosted outside the Stack Overflow websites.
\textbf{We do not consider the availability of the links that reference to the resources hosted within the Stack Overflow websites} because these links are maintained by Stack Overflow.
For example, the Stack Overflow moderators are aware of the deleted questions and have taken some actions, e.g., displaying the questions that are similar to the deleted question \cite{page_not_found_SO}.

We use the SOTorrent dataset\footnote{https://zenodo.org/record/3255045\#.XYWaMyh3iUk} \cite{baltes2018sotorrent, baltes2019sotorrent} to obtain the links in the \textbf{text} of posts (i.e., questions and answers) and comments on Stack Overflow.
SOTorrent dataset is based on the official Stack Overflow data dump that hosts the website data from Jul. 31, 2008 to Jun. 2, 2019.
Baltes et al. extracted and identified the text blocks and code blocks from all versions of posts from the Stack Overflow data dump Table \texttt{PostHistory} and stored these blocks into Table \texttt{PostBlockVersion} \cite{baltes2018sotorrent, baltes2019sotorrent}.
Baltes et al. collected the links in \textbf{the text blocks and the code blocks} in all versions of Stack Overflow posts using a regular expression and then stored these links into Table \texttt{PostVersionUrl}.
They also extracted the links in Stack Overflow comments with a regular expression and then store these links into Table \texttt{CommentUrl}.
We encourage readers to read their work for the full details of the data collection process of the SOTorrent dataset.

We extract the links \textbf{in the text of posts and comments} from the SOTorrent table \texttt{PostVersionUrl} and Table \texttt{CommentUrl}.
Links in the text of posts and comments are used to reference to resources.
We \textbf{do not consider the links in the code blocks} because some of these links, e.g., XML schema URIs, are not used for sharing knowledge.
We identify who shared the link and when the link was shared from Table \texttt{PostHistory} for Stack Overflow posts and Table \texttt{Comments} for comments.
We identify the links that reference to the resources hosted outside the Stack Overflow websites using their root domain as we mentioned above.
As a result, we finally obtain 12,446,901 links in Stack Overflow history, and 11,926,354 of them are in the latest version of Stack Overflow posts and comments (i.e., 520,574 links are not shared currently).

\subsection{Link Availability Test}\label{setup_linktest}

We perform the link availability test using Scrapy\footnote{https://scrapy.org/}.
Scrapy is an open-source web crawling and web scraping framework.
To identify the broken links, we obtain the HTTP response status code (i.e., response code) that is returned to the request for the resource referenced by the link.
To avoid the IP being banned from the website, we obtain a list of proxies from Free Proxy List\footnote{https://free-proxy-list.net/\#list} and make different requests using different proxies.
For the same website, we set a 15 seconds delay for different requests.
To reduce the bandwidth requirement, we only request the header of the response.
To mitigate the intermittent behavior, we perform one link availability test using an elastic compute service located in Virginia, U.S. in Dec 2019.
For the links that are not responded with 2xx response code, in Jan 2020, we perform another link availability test using an elastic compute service located in Singapore.
We identify the broken links that are not responded with 2xx response code in both trials.

\section{Preliminary Studies on the Broken Links on Stack Overflow}\label{s_pre}
In this section, we present a series of preliminary studies related to the broken links on Stack Overflow, including the prevalence of broken links, the response code of the broken links, and the broken links that were posted per month.

\subsection{Prevalence of Broken Links on Stack Overflow}\label{s_pre_prevalence}

\textbf{14.2\% (i.e., 1,687,995) of the links on Stack Overflow are broken links.
13.5\% (i.e., 2,156,095) of the posts and comments with links have broken links.
10.8\% (i.e., 2,493,328) of the occurrences of the links are broken links.}
Such a large proportion of broken links would downgrade the overall quality of the Stack Overflow.
However, it is still unclear what are the characteristics of the broken links on Stack Overflow.
This motivates us the further investigate the broken links on Stack Overflow.
By doing so, we could shed lights for future directions and provide recommendations for practitioners and researchers to address the broken links issue.

\subsection{Response Code of the Broken Links on Stack Overflow}\label{s_pre_responsecode}

To have a basic understanding of the broken links on Stack Overflow, we investigate the response code of the broken links.
To do so, we group the broken links according to the response code and count their corresponding numbers.

\begin{table*}[!htb]
    \centering
    \caption{Top 10 response codes returned for the broken links.
    This table shows that the 404 error is the main error code for the broken links on Stack Overflow.
    \# = number of broken links returning the corresponding status code.
    \% = percentage of broken links returning the corresponding status code.
    }\label{table_allcode}
    \begin{tabular}{|c|l|l|r|r|}
    \hline
        & HTTP Status Code & Explanation                                        & \#      & \%     \\ \hline
    1  & 404              & The requested resource could not be found currently.                               & 856,017 & 50.7\% \\ \hline
    2  & 403              & User not having the necessary permissions for the resource.                           & 220,261 & 13.0\% \\ \hline
    3  & DNSLookupError   & DNS lookup failed.                                                                 & 213,684 & 12.7\% \\ \hline
    4  & TCPTimedOutError & TCP connection timed out.                                                          & 66,940  & 4.0\%  \\ \hline
    5  & 405              & The request method is not supported for the requested resource.                      & 46,648  & 2.8\%  \\ \hline
    6  & 503              & The server cannot handle the request.                                              & 43,515  & 2.6\%  \\ \hline
    
    7  & 500              & The server failed to fulfill the request.                                          & 30,873  & 1.8\%  \\ \hline
    8  & 400              & The server cannot or will not process the request due to an apparent client error. & 19,030  & 1.1\%  \\ \hline
    9  & 401              & Authentication is required and has failed or has not yet been provided.            & 12,230  & 0.7\%  \\ \hline
    10 & 410              & The resource requested is no longer available and will not be available again.     & 10,148  & 0.6\%  \\ \hline
    \end{tabular}
\end{table*}

Table \ref{table_allcode} presents the top 10 response code of the broken links on Stack Overflow \cite{response_code}.
\textbf{50.7\% (i.e., 856,017) of the broken links are responded with the 404 response code.}
The 404 response code corresponds to a resource \textit{Not Found}; however, it does not indicate whether unavailability is temporary or permanent.
In contrast, the 410 response code explicits that the resource is likely to be permanently removed \cite{response_code}.
On Stack Overflow, only 0.6\% of the broken links are responded with the 410 response code.
We encourage the Stack Overflow moderators to remove the broken links that are responded with the 410 response code.

The 403 response code is another common response code among the broken links on Stack Overflow.
This response code indicates that the access to the resource requires authentication.
However, external visitors do not have the access.
Following the 404 and the 403 response code, DNS Lookup Error is the third and TCP Timed Out Error is the fourth most common status code for broken links.
One possible reason is that the website servers of these broken links fail to provide any response.

\subsection{Trendlines of Broken Links on Stack Overflow}\label{s_pre_trendline}

Here, we would like to investigate whether the links that were posted earlier on Stack Overflow are more likely to be broken.
By doing so, we can better understand whether the broken links on Stack Overflow is time-related.

\begin{figure}[!htb]
    \centering
    \includegraphics[width = \linewidth]{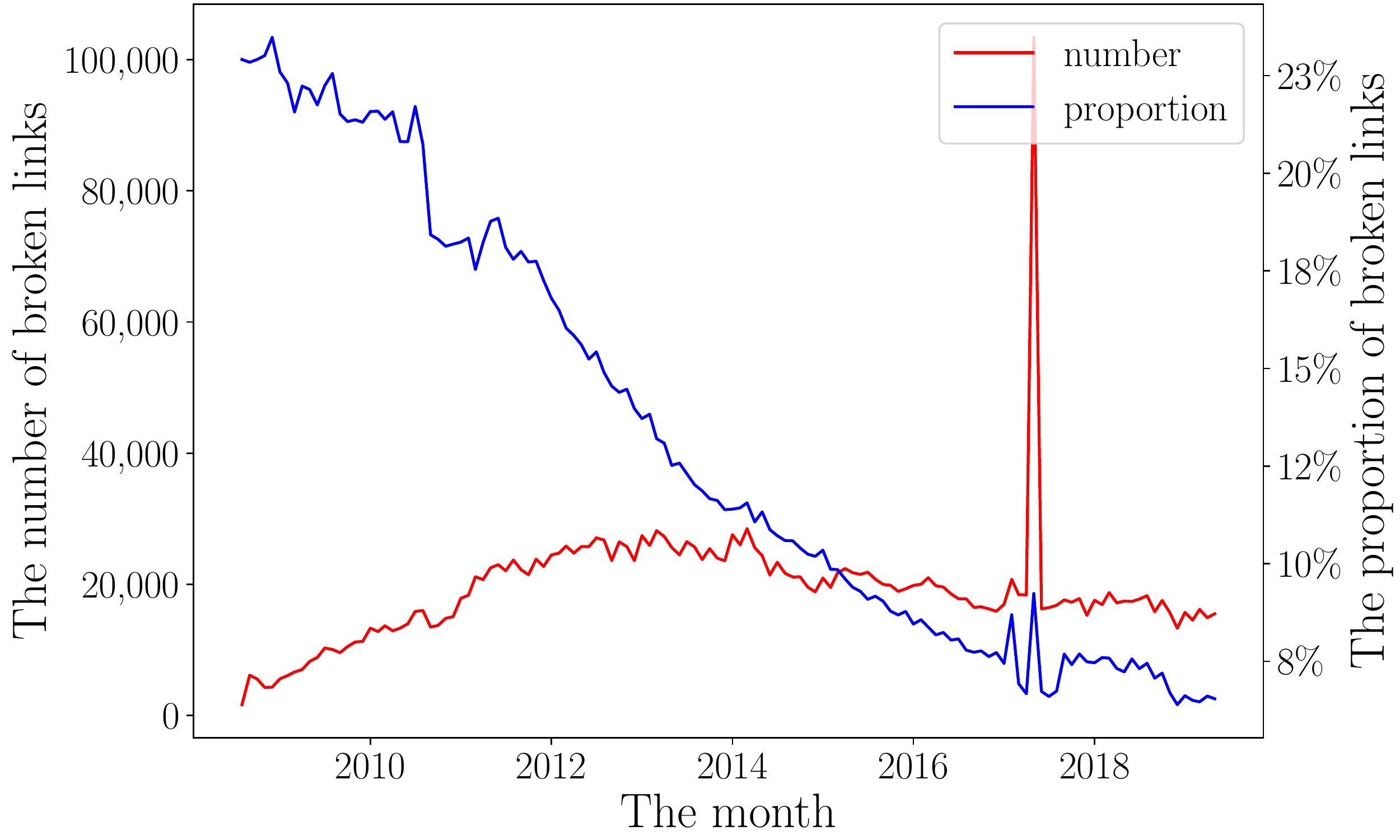}
    \caption{The numbers of broken links and the proportions of broken links among the links that were posted per month.
    This figure shows that the links that were posted earlier are more likely to be broken.}\label{figure_numbertrend}
\end{figure}

Figure \ref{figure_numbertrend} shows the numbers of broken links and the proportions of broken links among the links that were posted per month.
The mean and median number of broken links included in Stack Overflow posts in a month is 19,033 and 18,751 respectively.
The proportions of broken links among the links that were posted per month and the month the links are posted are significantly correlated with Pearson's correlation coefficient = -0.97 (p-value $<$ 0.05). 22.9\% of the links that were posted in Aug 2008 (the month when the Stack Overflow website was established) are broken.
\textbf{Links that were posted earlier are more likely to be broken.}
This indicates that broken links on Stack Overflow are time-related.

Stack Overflow posts made in May 2017 have a total of 103,411 broken links (the spike in Figure 2).
The number of broken links posted in May 2017 is 5.43 times greater than the average number of broken links that were posted per month.
We manually check the links that were posted in May 2017 and observe that 78.8\% (i.e., 882,362) of the links were posted by the URL Rewriter Bot.
As is indicated in \lstinline{meta.stackexchange.com} (i.e., the website for the meta-discussion of the Stack Exchange family of Q\&A websites), the URL Rewriter Bot is used by Stack Overflow to update the schema of the links, i.e., replacing HTTP with HTTPS for security and privacy concern without checking their validity\footnote{https://meta.stackexchange.com/q/291947/}$^,$\footnote{https://meta.stackoverflow.com/q/345012/}$^,$\footnote{https://nickcraver.com/blog/2013/04/23/stackoverflow-com-the-road-to-ssl/}.
However, 87,086 (i.e., 84.2\% of the broken links posted in May 2017) broken links were posted by the URL Rewriter Bot.
One possible reason is that widely applying the effective approach to resolve the link security problem, i.e., replacing HTTP with HTTPS, is much simpler than widely maintaining the broken links on Stack Overflow.

\section{Findings}\label{findings}

In this section, we present the results of our empirical study that answer the four research questions related to the broken links on Stack Overflow.
More specifically, we analyze the intended roles of the broken links in Stack Overflow questions, answers, and comments.
We investigate the impact of the broken links on the crowdsourced knowledge on Stack Overflow.
We also characterize the tags that are associated with broken links and the websites that are referenced by broken links.

\subsection{What are the Intended Roles of the Broken Links on Stack Overflow?}\label{findings_role}

\vspace{0.1cm}\noindent\textbf{Motivation:}
In Section \ref{s_pre_prevalence}, we have identified that broken links are prevalent on Stack Overflow.
Previous work observed different roles of the links on Stack Overflow \cite{Ye2017}.
However, it is still unclear what the \textit{intended} roles of the broken links in knowledge dissemination are.

\vspace{0.1cm}\noindent\textbf{Approach:}
To analyze the intended roles of the broken links in knowledge sharing, we follow Ye et al.'s work, where they analyzed the roles of sharing internal links on Stack Overflow \cite{Ye2017}.
Ye et al. observed there are five general roles of link sharing in Stack Overflow, i.e., 1) reference information for problem-solving, 2) reference existing answers, 3) reference visited but not helpful web pages, 4) recommend related information, and 5) others.
We randomly sampled a statistically representative sample of 384 posts and comments with broken links from questions, answers, and comments, respectively (i.e., 1,152 posts and comments broken links in total), using a 95\% confidence level with a 5\% confidence interval.
To label the intended roles of the broken links, we manually performed a lightweight open coding process to check the discussion context where the broken links are referenced.
This process involves 3 phases and is performed by the first two authors of this paper:

\begin{itemize}[wide=0pt]
    \item Phase I: We randomly selected 100 broken links from the sampled 1,152 broken links.
    The first two authors used the coding schema used by Ye et al.'s work to categorize the selected 100 broken links collaboratively \cite{Ye2017}.
    During this phase, the coding schema of the intended roles of the broken links on Stack Overflow was revised and refined.
    We performed the refinement because we observed that the intended role of many broken links is referencing information for problem-solving, we divided the intended role of ``reference information for problem-solving'' into two intended roles, i.e., providing working examples and providing supporting information.
    Table \ref{table_purpose} shows all the intended roles of broken links that we found.
    \item Phase II: The first two authors applied the resulting coding schema of Phase I to categorize the remaining 1,052 broken links independently.
    They were instructed to take notes regarding the deficiency and ambiguity of the coding schema for categorizing certain broken links.
    The inter-rater agreement (Cohen's kappa) of this stage is 0.69, indicating that the agreement level is substantial \cite{viera2005understanding}.
    \item Phase III: The first two authors discussed the coding results obtained in Phase 2 to resolve the disagreements.
    The first two authors revised the coding schema to resolve schema deficiencies and ambiguities.
    For example, for the broken link in an answer\footnote{https://stackoverflow.com/q/18369005/} to whether there is a ready Ajax extender or a JQuery functionality to implement search textbox,
    
    \vspace{0.1cm} \hangindent 1.5em\textit{For the full code check out the following post: http://www.simplygoodcode.com/2013/08/placing-text-and-controls-inside-text.html\footnote{http://www.simplygoodcode.com/2013/08/placing-text-and-controls-inside-text.html}}

    \vspace{0.1cm}
    \noindent\normalsize
    The first author considered the intended role of this broken link is to provide supporting information because the post owner encourages viewers to check out the \textit{post} for more details.
    The second author considered the intended role of this broken link is to show code examples as ``the full code can be found in the broken link''.
    We finally consider the intended role of this broken link is to provide supporting information as the broken link is a personal blog that records the approaches to solve programming related problems.
    At the end of Stage 3, we obtained the final coding schema and the final coding results of the sampled 1,152 broken links.
\end{itemize}

\begin{savenotes}
    \begin{table*}\centering
        \caption{Intended roles for broken links.
        This table shows that most of the broken links in questions are used to show examples, and most of the broken links in answers are used to provide supporting information on users' claims.}\label{table_purpose}
        \begin{tabular}{|m{1.3cm}<{\raggedright}|m{4cm}<{\raggedright}|m{5.5cm}<{\raggedright}|l|l|l|}
        \hline
        \textbf{Intended Role}       & \textbf{Definition}                                                                                                                                                                                                                                                                                      & \textbf{Example}                                                                                                                                                                                                                                                                                                                   & \textbf{\% Questions} & \textbf{\% Answers} & \textbf{\% Comments} \\ \hline
        Working Example        & Provide working examples, e.g., code snippets.                                                                                                                                                                                                                              & if you go to \underline{wp.matthewwood.me\footnote{http://wp.matthewwood.me/}} and click through the links you will see what I mean ... \footnote{https://stackoverflow.com/q/31938291/}                                 & 65\%                  & 22\%                & 49\%                 \\ \hline
        Supporting Information & Explain a certain concept, approach of (sub)step to solve the questions, background knowledge, or the link sharer's claim.                                                                                                                                                  & However I followed the tutorials where I create a plugin, feature ... \underline{http://mattiasholmqvist.se/2010/03/bui}-\underline{lding-with-tycho-part-2-rcp-applications/\footnote{http://mattiasholmqvist.se/2010/03/building-with-tycho-part-2-rcp-applications/}} seems a bit out of date.\footnote{https://stackoverflow.com/q/18550820/}                                          & 22\%                  & 70\%                & 44\%                 \\ \hline
        Existing Answers       & Reference existing answers in a Q\&A websites.                                                                                                                                                                             & Interesting read on the same topic - \underline{quora.com/...\footnote{https://www.quora.com/Java-When-we-concatenate-two-strings-using-the-+-operator-will-the-resulting-string-be-stored-in-the-string-literal-pool-or-not?share=1}}...\footnote{https://stackoverflow.com/q/44050772/}                                                                                                                                                                                                                                                                      & 0\%                   & 0\%                 & 1\%                  \\ \hline
        Visited Webpages       & Reference visited ``search and research'' web pages that cannot solve the problem. & Things I've already read ... \underline{CMake bug report 0013765\footnote{http://public.kitware.com/Bug/bug\_revision\_view\_page.php?-rev\_id=958}} – this\footnote{https://stackoverflow.com/q/15159722/} & 13\%                  & 0\%                 & 2\%                  \\ \hline
        Related Information    & Recommend related information that does not directly answer the question.                                                                           & I think so, more details please reference \underline{this article\footnote{https://msdn.microsoft.com/en-us/windows/uwp/globalizin-g/put-ui-strings-into-resources}}\footnote{https://stackoverflow.com/q/40120304/}    & 0\%                   & 8\%                 & 3\%                  \\ \hline
        Others                 & Suggest how to good asks.                                                                                                                          & And what else? Could you add some more information. Check this metaSO question and \underline{Jon Skeet: Coding Blog\footnote{http://msmvps.com/blogs/jon\_skeet/archive/2009/02/17/an-swering-technical-questions-helpfully.aspx}} on how to give a correct answer.\footnote{https://stackoverflow.com/q/12838984/}                                                                                                                                            & 0\%                   & 0\%                 & 1\%                  \\ \hline
        \end{tabular}
    \end{table*}
\end{savenotes}

\begin{table*}\centering
    \caption{Prevalence of the broken links in Stack Overflow questions, answers, and comments.
    This table shows that questions and comments have higher proportions of broken links, and answers contribute the largest number of broken links.
    }\label{table_prevalence}
    \begin{tabular}{|c|r|r|r|r|r|r|}
    \hline
     & \textbf{\# Posts} & \textbf{\% Posts} & \textbf{\# Links} & \textbf{\% Links} & \textbf{\# Occurrences} & \textbf{\% Occurrences} \\ \hline
    Questions & 620,837  & 16.8\%            & 635,062  & 14.4\%            & 752,414        & 13.9\%                  \\ \hline
    Answers   & 905,964  & 11.0\%            & 670,145  & 11.0\%            & 1,061,838      & 8.5\%                   \\ \hline
    Comments  & 641,448  & 13.7\%            & 556,417  & 18.5\%            & 679,076        & 13.2\%                  \\ \hline
    \end{tabular}
\end{table*}

\vspace{0.1cm}\noindent\textbf{Results:}
Table \ref{table_prevalence} shows the prevalence of the broken links in all types of Stack Overflow posts and comments.
We find that the proportions of questions and comments that have broken links among all questions and comments are higher than answers -- the ratios are 1:1.52 and 1:1.24 respectively.
The proportions of broken links among all links in questions and comments are higher than answers -- the ratios are 1:1.31 and 1:1.68 respectively.
The proportions of the occurrences of broken links among the occurrences of all links in questions and comments are higher than answers -- the ratios are 1:1.63 and 1:1.55 respectively.
The above indicates that \textbf{questions and comments have higher proportions of broken links than answers.}

Moreover, the number of answers that have broken links is higher than questions and comments –- the ratios are 1:1.46 and 1:1.41 respectively.
The number of broken links in answers is higher than questions and comments –- the ratios are 1:1.06 and 1:1.20 respectively.
The number of the occurrences of broken links in answers is higher than questions and comments –- the ratios are 1:1.41 and 1:1.56 respectively.
This shows that \textbf{answers contribute more to the absolute numbers of broken links than questions and comments.}
10.1\% (i.e., 297,305) of the links in accepted answers are broken links.
10.7\% (i.e., 338,705) of the accepted answers have broken links.
44.4\% (i.e., 297,305) of the broken links in answers are posted in accepted answers.
This shows that \textbf{broken links are common in the accepted answers.
The answers with broken links can solve questioners' problems when the links were posted, i.e., before the links were broken.}

Table \ref{table_purpose} shows the intended roles of the broken links in Stack Overflow questions, answers, and comments, respectively.
\textbf{We observe that 65\% of the broken links in our sampled questions are used to show examples, e.g., the demos of the tasks and the code examples written by the post owners.}
Questioners explain the tasks with these links when they post their questions.
These links may reference to test cases, demos, or the development versions of a software.
After the problem was resolved, the questioners removed the example from the link.
This practice cause the link to be broken.
However, without these links, the following viewers cannot tell whether the question is similar or even identical to the problems they are facing with.
Because the following viewers cannot benefit from the questions, the broken links in questions can lead to the questions with broken links to be useless.
For example, in a comment to a question\footnote{https://stackoverflow.com/q/7589262/}:

\vspace{0.1cm} \hangindent 1.5em\textit{Do you still have that code? If so, please edit it in. Your Dropbox link is dead so the question is useless now.
  }\vspace{0.1cm}
  
\noindent\normalsize
The broken link\footnote{http://dl.dropbox.com/u/3085200/canvasTest/index.html} is the test page to show an example of the questioner's problem.
Viewers cannot understand the question without the example hosted in the broken link.
We suggest that users should not remove the examples in links.

\textbf{In our sampled answers, 70\% of the broken links are used to provide supporting information, e.g., a certain concept and a step to solve a problem}.
Such information is provided by the Stack Overflow communities to solve programming-related questions.
For example, in a comment to an accepted answer\footnote{https://stackoverflow.com/q/15488527/15489656}:

\vspace{0.1cm} \hangindent 1.5em\textit{The link is dead. Could you fix it? I'm very much interested in a solution to this problem.
  }\vspace{0.1cm}
  
\noindent\normalsize
The viewer complains that he cannot fully understand the answer with the broken link.
The broken link\footnote{http://home.roadrunner.com/~hinnant/stack\_alloc.html} explains the substep to solve the problem.
This leads to the answers useless and damage the reputation of the Stack Overflow.

\textbf{94\% of the broken links in our sampled comments can be used in problem-solving.}
Table \ref{table_purpose} shows that the broken links in comments can be used in problem solving to provide working examples (i.e., 49\%), supporting information (i.e., 42\%), existing answers (i.e., 1\%), and visited but not helpful knowledge (i.e., 2\%), etc.
This finding is consistent with Zhang et al.'s work \cite{zhang2019comments}, where they observed that comments on Stack Overflow can be leveraged to improve the quality of the associated answers \cite{zhang2019comments}.
However, we observe that broken links are common in comments.
We suggest the Stack Overflow moderators should pay attention to the maintenance of comments as well.

\rqbox{Questions and comments have higher proportions of broken links than answers.
Stack Overflow answers contribute broken links the most compared with questions and comments.
65\% of the broken links in our sampled questions are used to show examples, e.g., code examples.
70\% of the broken links in our sampled answers are used to provide supporting information, e.g., explaining a certain concept and describing a step to solve the problem.
}

\subsection{What are the Impacts of the Broken Links on Stack Overflow?}\label{findings_mechanism}

\vspace{0.1cm}\noindent\textbf{Motivation:}
In Section \ref{ss_rw_mining_so}, we introduce that the registered viewers can vote up the crowdsourced knowledge (i.e., questions and answers) that are useful to them, i.e., can provide values to the developers who encounter similar problems that are already asked on Stack Overflow \cite{vote_up, why_vote}.
Users can increase their reputations by posting high-quality questions and answers.
However, it is still unclear how broken links impact the crowdsourced knowledge on Stack Overflow.
More specifically, we would like to characterize whether the posts with broken links can help users in programming related problem solving as other posts, i.e., whether the posts with broken links receive fewer votes after the links become broken?

\vspace{0.1cm}\noindent\textbf{Approach:}
To investigate the impacts of broken links on Stack Overflow, we extract the vote scores of posts and the view count of questions from the Stack Overflow data dump Table \texttt{Posts}.
We extract the date of each vote from the Stack Overflow data dump Table \texttt{Votes}.

We first analyze whether viewers would vote on the posts without broken links more than the posts with broken links before the links were broken and after the links were broken.
To estimate the votes that were received before the link becomes broken, we collect the vote to the posts with broken links in the first 30 days after being posted.
However, the links could be broken in the first 30 days after being posted.
To get the proportion of broken links that have been broken in the first 30 days, we take the links posted on Stack Overflow 30 days before our paper is written as an example.
To do so, we collect the posts and comments with links (i.e., marked with <a>…</a>) that were posted from Aug. 26, 2020, to Sept. 26, 2020, from Stack Exchange Data Explorer\footnote{https://data.stackexchange.com/}.
Stack Exchange Data Explorer is an open-source tool for running queries against the public data that is similar to the data in Stack Exchange data dumps.
As a result, we obtain 89,632 posts and comments that were posted with external links from Aug. 26, 2020, to Sept. 26, 2020.
Then, we randomly sampled a statistically representative sample of 661 posts and comments with links from the 89,632 posts and comments, using a 99\% confidence level with a 5\% confidence interval.
We visit the sampled posts and comments and randomly click one of the links in the posts and comments.
As a result, we find 4 broken links, i.e., 0.6\% of the links that were posted in 20 days were broken links.
We multiply the number of links posted each month by this proportion to estimate the number of links that were broken within the first 30 days after being posted.
As a result, only 6\% (i.e., 139,427) of broken links have been broken in the first 30 days.
This gives us high confidence in using the vote to the posts with broken links in the first 30 days after being posted to estimate the votes that are received before the link becomes broken.
We also collect the vote to the posts without broken links in the first 30 days after being posted to see whether there is any difference between the posts with broken links and the posts without broken links before the links become broken.

To estimate the votes that are received after the link becomes broken, we collect the vote to the posts with broken links in the last 30 days before the collection of the dataset.
However, the links could not be broken in the last 30 days before the collection of the dataset.
As indicated in Section \ref{setup_linktest}, we perform two link availability tests to mitigate the intermittent behavior.
To get the proportion of broken links that have not been broken in the last 30 days, we count the number the broken links that are identified in the link availability test that is performed in Jan 2020 but not in Dec 2019.
As a result, we observe 62,441 broken links that are only observed in Jan 2020.
This indicates that 97.5\% (i.e., 2,430,887) of the broken links have been broken links in the last 30 days before the collection of the dataset.
This gives us high confidence in using the vote to the posts with broken links in the last 30 days before the collection of the dataset to estimate the votes that are received after the link becomes broken.
We also collect the vote to the posts without broken links before the collection of the dataset to see whether there is any difference between the posts with broken links and the posts without broken links after the links become broken.

We then characterize whether viewers are more likely to vote on the posts without broken links than the posts with broken links after browsing since the post of the links.
To do so, for Stack Overflow questions, we compare the vote scores of the questions with broken links with the questions without broken links in the same range of view counts.
For Stack Overflow answers, we exclude the questions that only have one answer and present the ranks of the answers with broken links among all answers to a specific question in terms of the vote scores.

To capture an overview of how often did \textbf{viewers notify the posts owners of the broken links}, we extract \textbf{the comments indicating the notification of broken links in posts}.
More specifically, we identify 2,727,969 comments to the posts with broken links.
To identify the comments indicating the notification of broken links in posts, we use the keywords that indicate broken links in previous work \cite{habibzadeh2013decay, fetterly2004large, koehler2004longitudinal, mccown2005availability, hennessey2013cross}, wordnet\footnote{http://wordnetweb.princeton.edu/perl/webwn}, and other online resources\footnote{https://en.wikiredia.com/wiki/Wikipedia:Link\_rot}.
For example, ``broken'', ``404'', ``unavailable'', ``rot'', ``inaccessible'', ``dead'', and so on, can represent the keyword ``broken'';
``url'', ``reference'', ``citation'', and so on, can represent the keyword ``link''.
Because there are various reasons why links are updated, e.g., updating the obsolete knowledge, we do not use the keyword ``update'' to identify the comments indicating the notification of broken links.
\textbf{Finally, we obtain a total of 27,261 comments indicating the notification of broken links in 25,443 posts.}
2,700,708 comments are not identified as the comments indicating the notification of broken links.
To check the precision of identifying the comments indicating the notification of broken links, we randomly sampled a statistically representative sample of 379 comments from 27,261 identified comments using a 95\% confidence level with a 5\% confidence interval.
We find that our heuristics achieves a precision score of 0.88 as 42 comments that are false positive (i.e., not the comments indicating the notification of broken links).
Similarly, to check the recall of identifying the comments indicating the notification of broken links, we randomly sampled a statistically representative sample of 379 comments from 2,700,708 comments using a 95\% confidence level with a 5\% confidence interval.
We find that there is no comment indicating the notification of broken links.

To mitigate the negative impact brought by broken links, Stack Overflow sets up community norms in their users' guides.
For example, Stack Overflow suggests questioners copy the code of the live example of the problem \cite{howtoask}.
Stack Overflow also suggests answerers quote the most relevant part of an important link \cite{howtoanswer}.
To capture an overview of how often did users follow the community norms when posting links, we manually label the 768 posts with broken links (i.e., 384 questions and 384 answers) that are sampled from Section 5.1.
More specifically, we manually label whether there is a code snippet or quotation box that quotes the content from the broken link or any links in the posts based on the context of the links.

To capture an overview of how often users removed the broken links, we compare the number of posts that do not have broken links currently with the number of posts that had broken links in history.

To analyze whether the use of broken links is related to the user reputation, we extract the id and the name of the users who posted the links from the Stack Overflow data dump Table \texttt{PostHistory}.
We extract the user reputations from the Stack Overflow data dump Table \texttt{Users}.
Then we check the proportion of broken links among the links posted by users with different reputation ranges.

Considering the impact of broken links on the crowdsourced knowledge on Stack Overflow, we are interested in whether the future viewers would encounter more broken links when browsing the posts with certain vote scores.
As registered viewers can vote up the posts that are useful to them, posts with higher vote scores indicate higher values to viewers.
Viewers would expect to receive more help from the posts with higher vote scores.
To do so, we check the proportion of broken links among the links in the questions and the answers with different vote score ranges.

\vspace{0.1cm}\noindent\textbf{Results:}

\begin{figure*}\centering
    \subfloat[The numbers of votes per post in the first 30 days since the sharing of links or the creation of the posts.
]{
        \includegraphics[width = 0.5\linewidth]{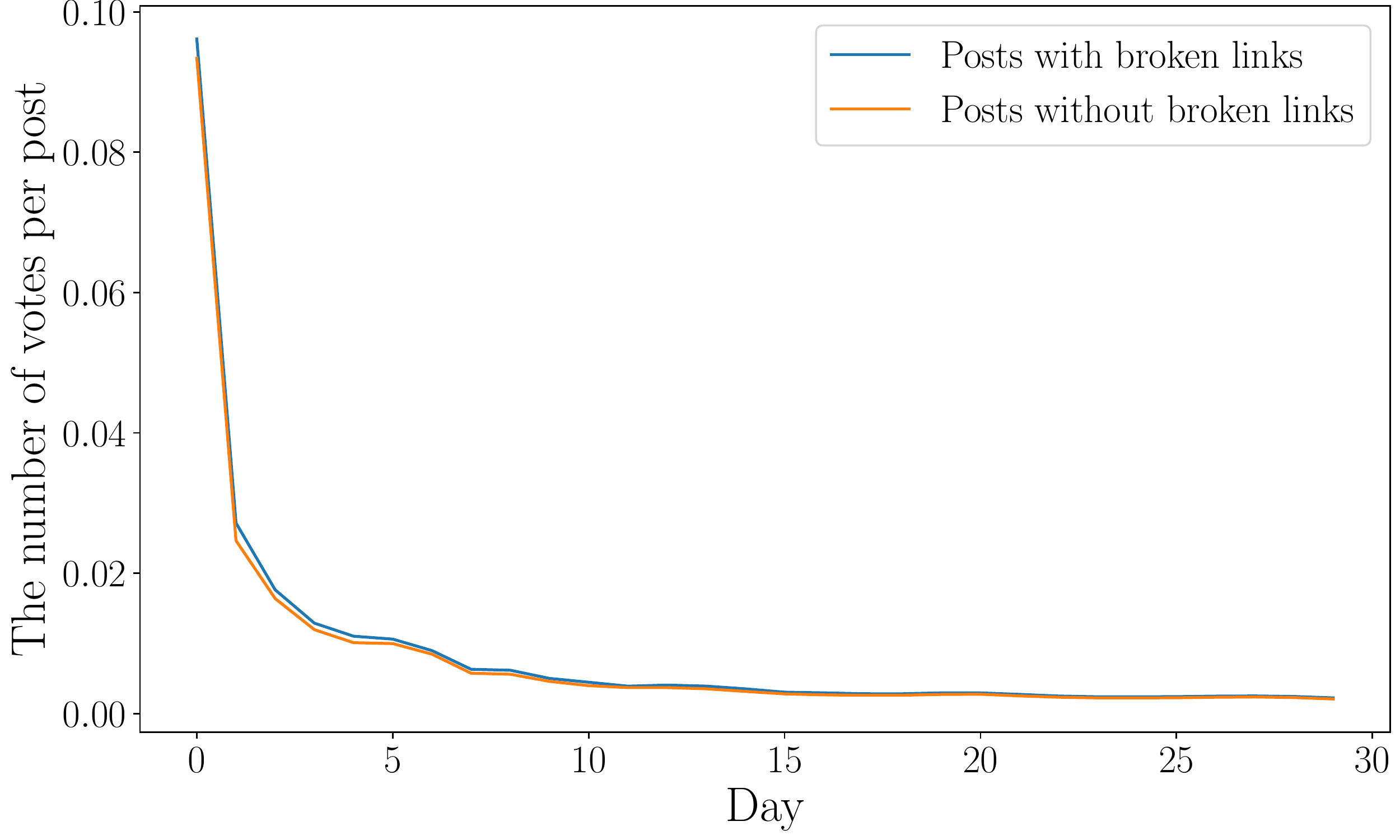}\label{figure_vote_before}
    }
    \subfloat[The numbers of votes per post in the last 30 before the collection of dataset.
]{
        \includegraphics[width = 0.5\linewidth]{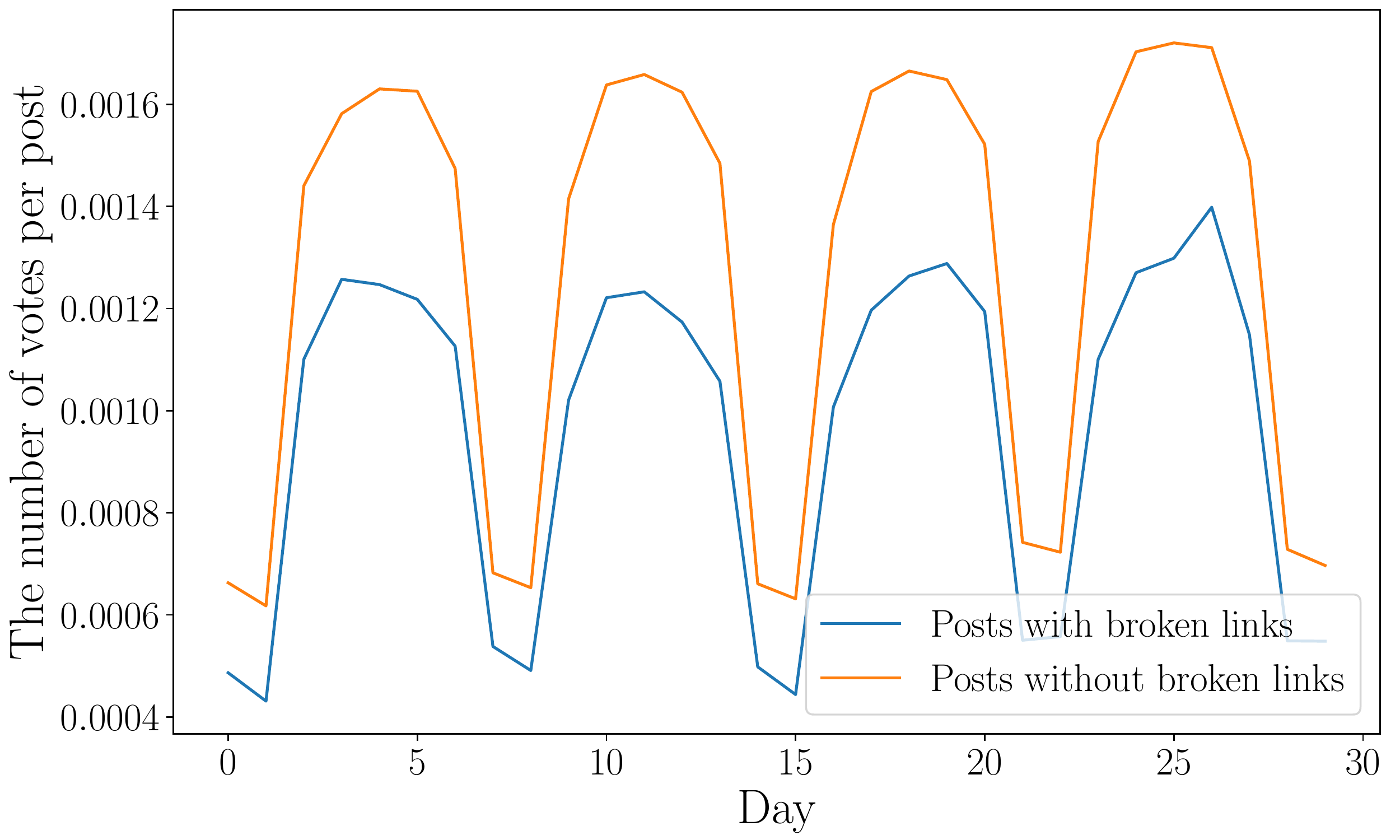}\label{figure_vote_last}
    }
    \caption{The vote scores of the posts with broken links and the posts without broken links before the links were broken and after the links were broken.
    These figures show that viewers would vote on the posts with broken links as the posts without broken links before the links were broken, but viewers would vote less on the posts with broken links than the posts without broken links after the links were broken.
    }
\end{figure*}

\textbf{Viewers would vote on the posts with broken links in the same way as the posts without broken links before the links were broken.} Figure \ref{figure_vote_before} shows the numbers of votes per post of the posts current with broken links and the posts current without broken links in the first 30 days after the creation of posts.
To check whether the differences in the number of votes per post in the first 30 days after the creation of posts are statistically significant between the posts current with broken links and the posts current without broken links, we perform a Mann Whitney test \cite{mann1947test}.
The null hypothesis is that there is no difference between the posts current with broken links and the posts current without broken links in terms of the number of votes per post in the first 30 days after the creation of posts.
As a result, there is no difference between the posts current with broken links and the posts current without broken links in terms of the number of votes per post in the first 30 days after the creation of posts (p-value > 0.05).
This indicates that the posts current with broken links could also solve viewers' problems before the links were broken as the posts without broken links.

\textbf{Viewers would vote less on the posts with broken links than the posts without broken links after the links were broken.}
Figure \ref{figure_vote_last} shows the numbers of votes per post of the posts with broken links and the posts without broken links in the last 30 before the collection of the dataset.
To check whether the differences between the posts with broken links and the posts without broken links are statistically significant in the number of votes per post in the last 30 days before the collection of dataset, we perform a Mann Whitney test \cite{mann1947test}.
As a result, the difference between posts with broken links and the posts without broken links in the number of votes per post in the last 30 days before the collection of the dataset is significant (p-value < 0.05).
We then calculate Cliff's delta to measure the effect size \cite{cliff2014ordinal}.
Cliff's delta is a non-parametric effect size measure that can evaluate the amount of difference between two groups.
It defines an absolute delta of less than 0.147, between 0.147 and 0.33, between 0.33 and 0.474 and above 0.474 as ``Negligible'', ``Small'', ``Medium'', ``Large'' effect size, respectively.
As a result, the difference between posts with broken links and the posts without broken links in the number of votes per post in the last 30 days before the collection of the dataset is large (Cliff's delta > 0.474).
This indicates that the posts with broken links are inferior to the posts without broken links in solving viewers' problems, and broken links in posts have a negative impact on the crowdsourced knowledge on Stack Overflow

\textbf{Only 5.8\% (i.e. 103,792) of the broken links were removed.}
5.6\% (i.e., 90,606) of the posts with broken links removed the broken links.
6.7\% (i.e., 164,404) of revisions that changed the links in posts removed broken links.
This shows that broken links attract limited attention on Stack Overflow.
\textbf{1.57\% (i.e., 25,443) of the posts with broken links are notified of the broken links with via one or more comments (i.e., notified posts).
14.3\% (i.e., 3,648) of the notified posts removed the broken links.}
The proportion of the notified posts that removed broken links among all notified posts is 2.47 times larger than the proportion of the posts that removed broken links among the posts with broken links.
This shows that when notified of the broken links, users are more likely to repair the broken links.
We suggest Stack Overflow could design a mechanism to notify broken links in posts.

21.1\% (i.e., 81) answers and 15.1\% (i.e., 58) questions with broken links follow the community norms, i.e., quote the content of any links in the posts into quotation boxes, code blocks, or code snippets.
This shows that \textbf{users commonly not follow the Stack Overflow community norms.}
10.2\% (i.e., 39) answers and 12.5\% (i.e., 48) questions with broken links quote the code of broken links.
5.5\% (i.e., 21) answers and 0.3\% (i.e., 1) questions with broken links quote the text of broken links.
This shows that users pay more attention to the permanent visit of their code.
5.5\% (i.e., 21) answers and 2.3\% (i.e., 9) questions also quote other links that are not broken links.
70.1\% (i.e., 60) answers and 84.5\% (i.e., 49) questions that follow the community norms quote the broken links.
Viewers cannot obtain the content of the broken links via a click but viewers can obtain the content of the broken links via the quotations from these links.
This also shows the importance of broken links, i.e., when users need to quote an important link, users would quote the broken links.
We suggest researchers design a tool to identify the links that need quotation.

\begin{figure*}\centering
    \subfloat[The ratios of the score of questions without broken links to the score of questions with broken links in different view count interval.]{
        \includegraphics[width = 0.5\textwidth]{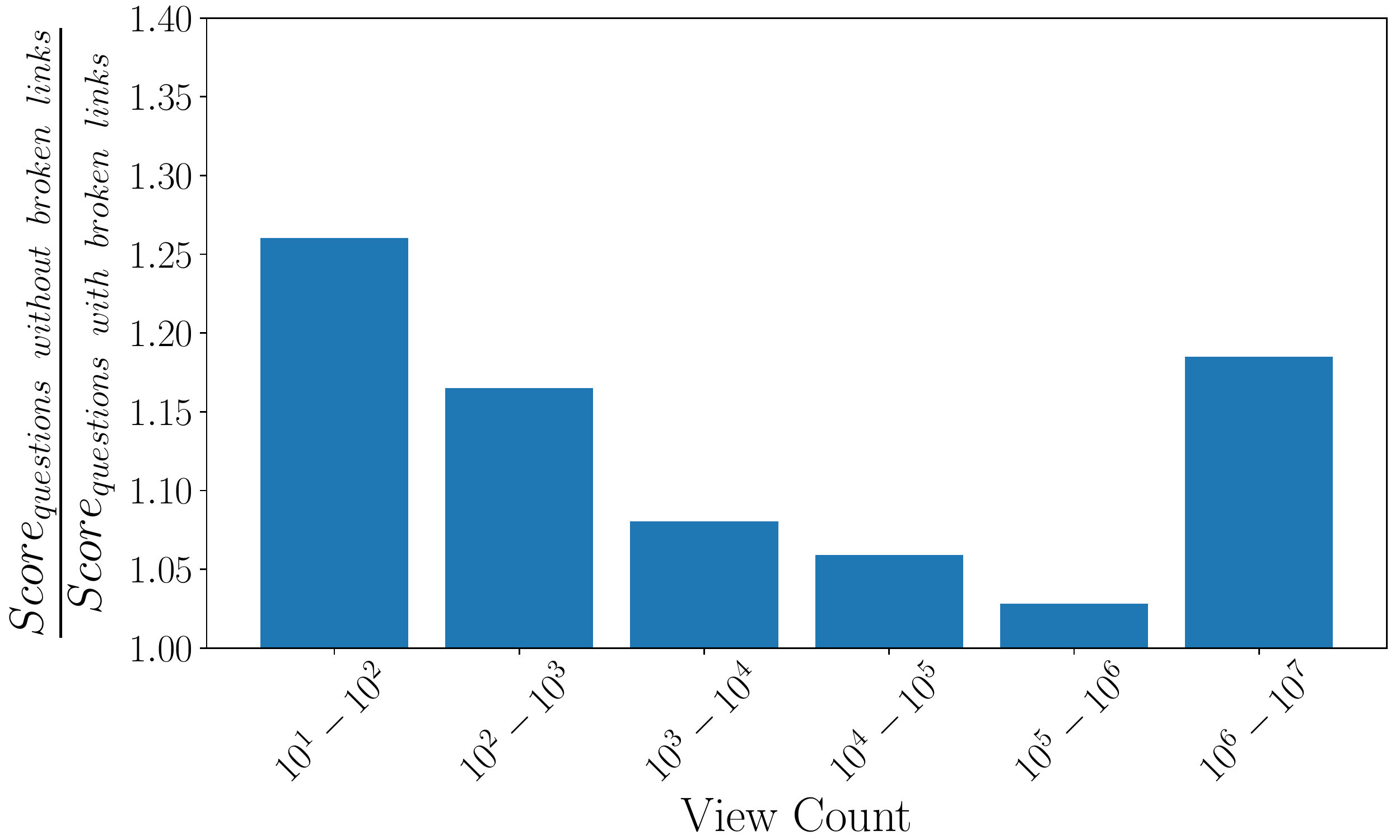}\label{figure_view_score}
    }
    \subfloat[The ranks of the answers with broken links among all answers to a specific question in terms of the vote scores.]{
        \includegraphics[width = 0.5\textwidth]{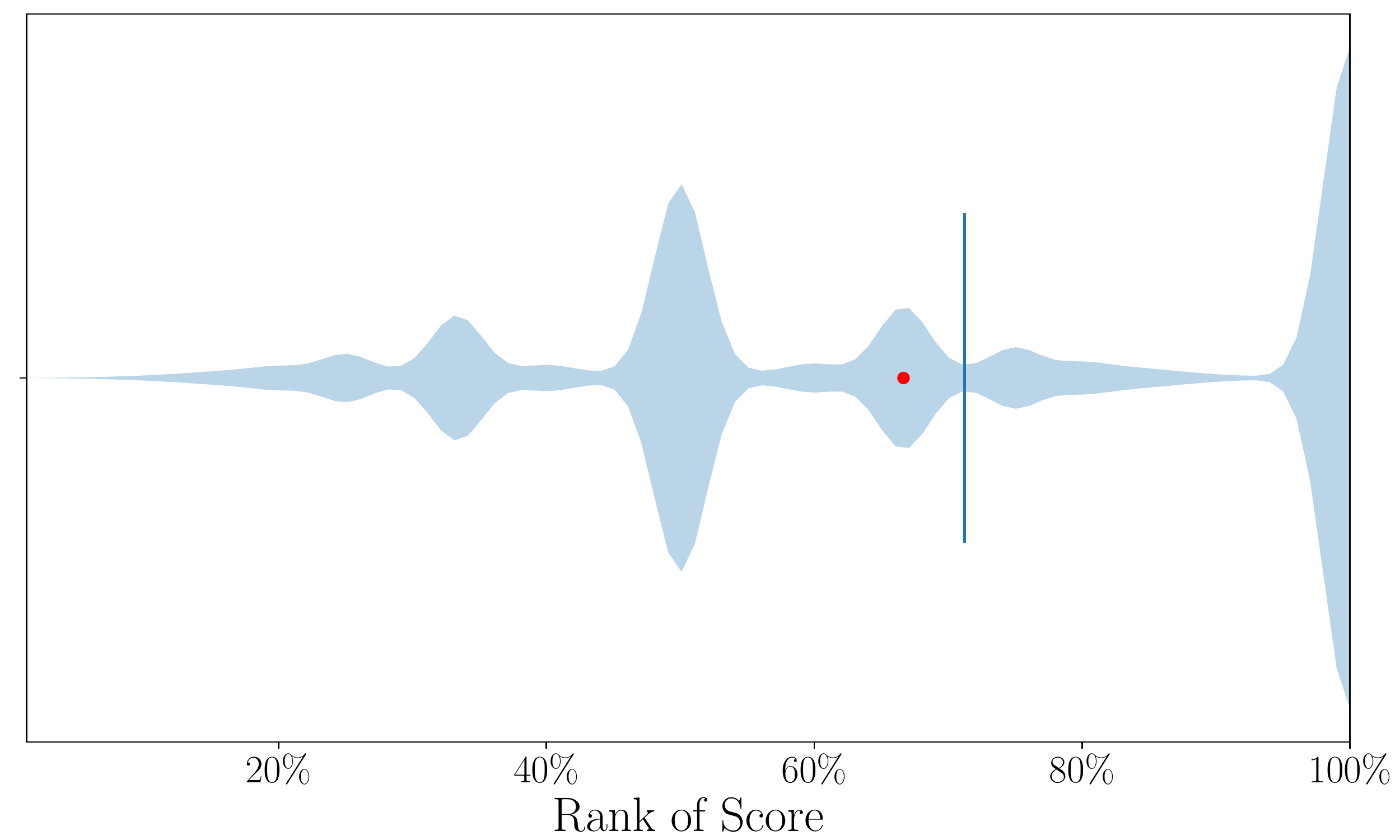}\label{figure_answer_score}
    }
    \caption{The vote scores on the questions and answers with or without broken links.
    These figures show that viewers are less likely to vote on posts with broken links.
    }
\end{figure*}

For Stack Overflow questions, Figure \ref{figure_view_score} shows the relations between the overall view counts and the vote scores.
To check whether the differences in the number of vote scores per view count are statistically significant between the questions with broken links and the questions without broken links, for the questions with the different ranges of view counts, we perform a Mann Whitney test \cite{mann1947test}.
The null hypothesis is that there is no difference between the questions with broken links and the questions without broken links in terms of the number of vote scores per view count in different ranges of view counts.
As a result, the difference between the questions with broken links and the questions without broken links is significant (p-values $<$ 0.05).
We then calculate Cliff's delta to measure the effect size \cite{cliff2014ordinal}.
As a result, the difference between the questions with broken links and the questions without broken links in terms of the number of vote scores per view count in different ranges of view counts is medium (Cliff's delta is between 0.33 and 0.474).
More specifically, we observe that \textbf{for the questions with the same range of view counts, questions without broken links have higher vote scores than the questions with broken links}.
When viewers browse the Stack Overflow website, they would encouter the questions with broken links and the questions without broken links.
However, questions with broken links are less likely to be voted after being browsed.
This shows that viewers waste effort in reading questions with broken links and finally vote on the questions without broken links to express the usefulness of the questions.
\textbf{The questions with broken links have fewer values to viewers than the questions without broken links in terms of the vote scores.}

Figure \ref{figure_answer_score} shows the ranks of the answers with broken links among all answers to a specific question in terms of the vote scores.
More specifically, we observe that only 16.1\% of the answers with broken links are ranked as the top 50\% answers to a specific question in terms of the vote scores.
40.5\% answers with broken links have the least vote scores.
\textbf{This shows that for the answers to the same questions, answers without broken links have higher vote scores than the answers with broken links.}
When viewers browse the answers to the questions, they vote on the answers without broken links.
This shows that \textbf{the answers with broken links have fewer values to viewers than the answers without broken links in terms of the vote scores.}

\begin{figure*}\centering
    \subfloat[Question]{
        \includegraphics[width = 0.5\textwidth]{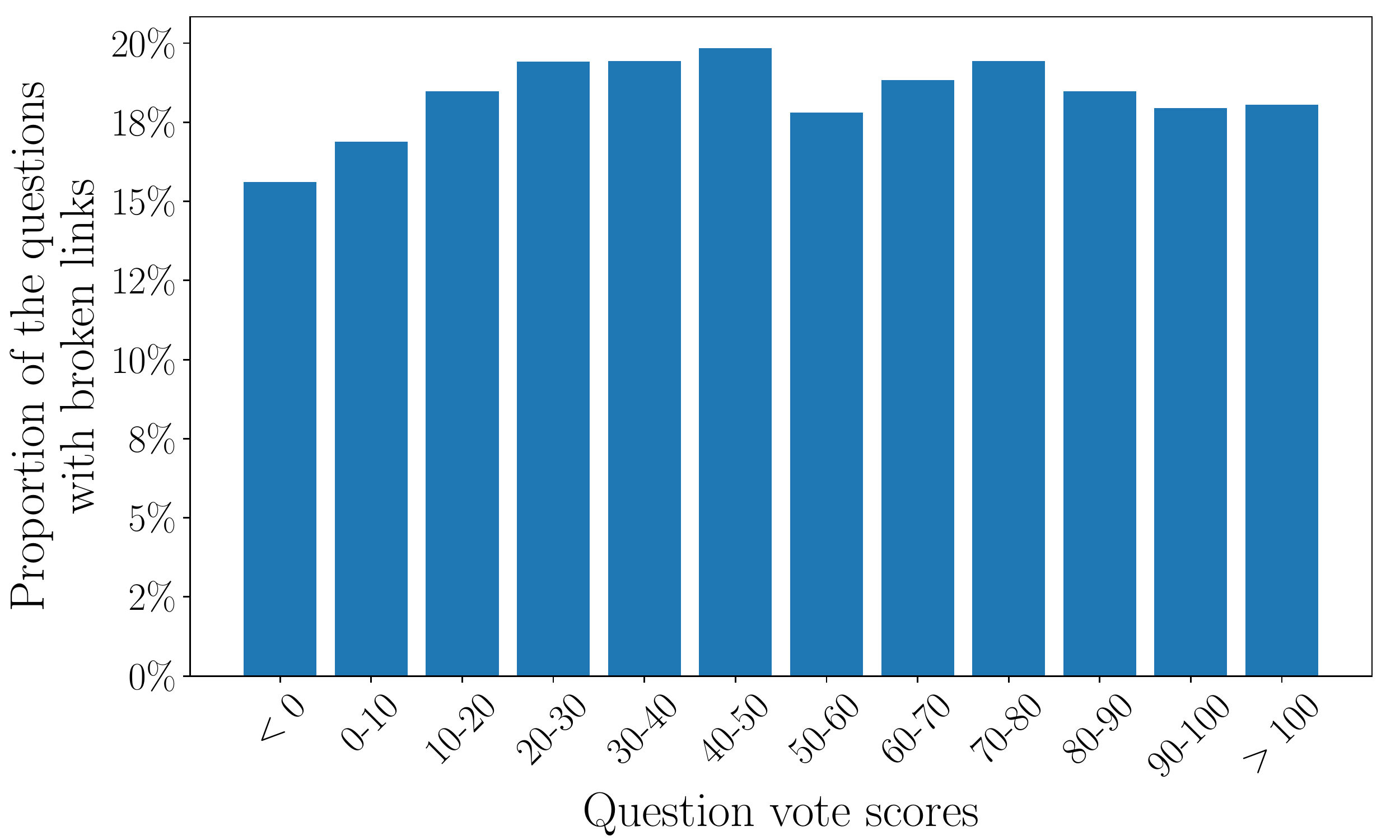}\label{figure_question_score}
    }
    \subfloat[Answer]{
        \includegraphics[width = 0.5\textwidth]{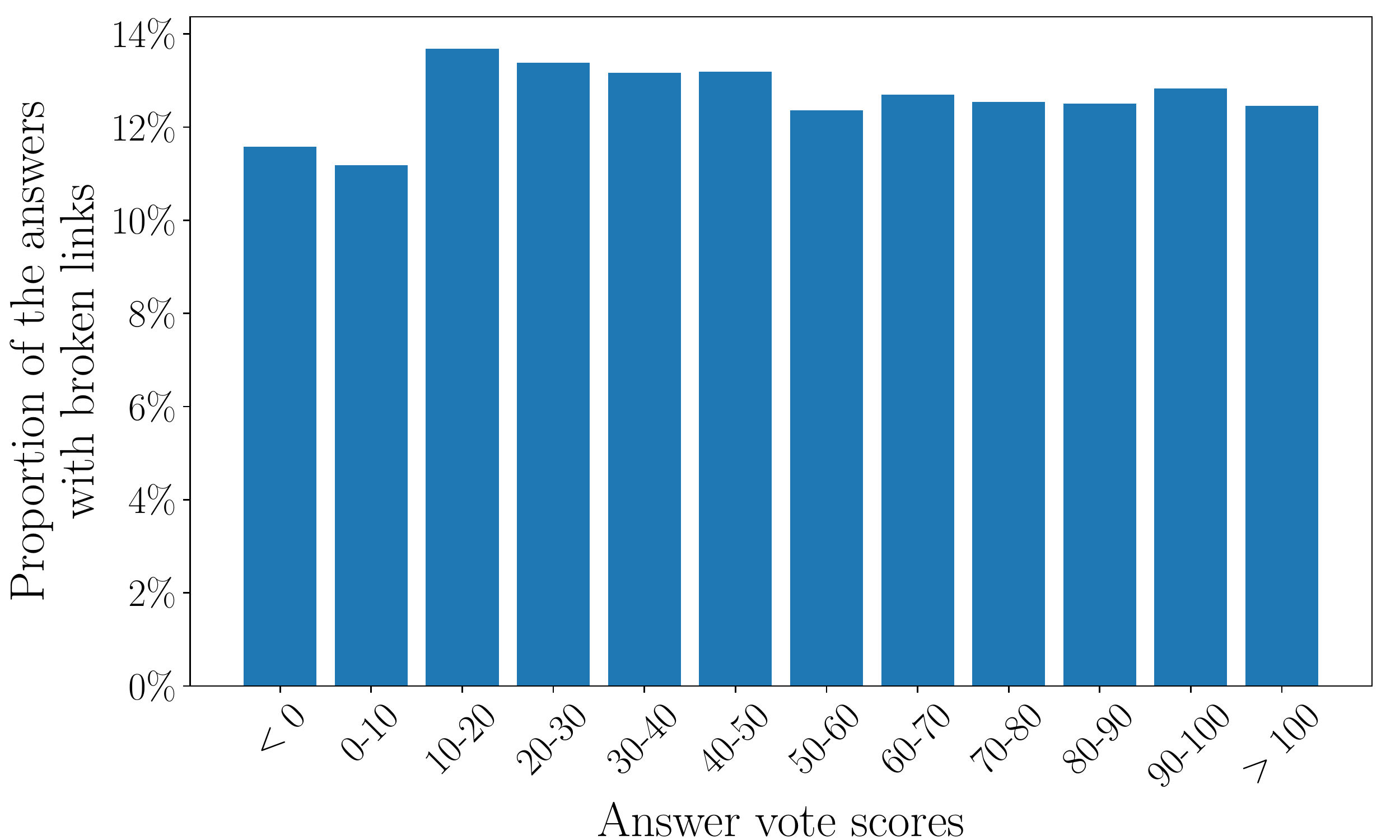}\label{figure_answer_score}
    }
    \caption{The proportions of the posts with broken links among the posts with links based on the vote scores.
    These figures show that broken links are common across questions and answers with different vote scores.
    }\label{figure_score}
\end{figure*}

\textbf{Viewers cannot fully rely on the vote scores to detect broken links, as broken links are common across posts with different vote scores}.
Figure \ref{figure_score} shows the distribution of the proportions of broken links among the links in the questions and the answers with different vote scores.
These proportions range from 15.6\% to 19.8\% in questions and 11.1\% and 13.6\% in answers.
This shows that although Stack Overflow viewers are less likely to vote on the posts with broken links, broken links are common across posts with different vote scores.
For questions, one possible reason is that the questions with broken links with higher view counts can have higher vote scores than the questions without broken links with lower view counts.
Similarly, one possible reason for answers is that the answers with broken links can have higher vote scores than the answers to another question without broken links.
When viewers browse the posts that have been voted by other viewers, it is common for them to encounter broken links.
We suggest Stack Overflow could detect the broken links and mark up the posts with broken links.

\begin{figure}\centering
    \includegraphics[width = 0.5\textwidth]{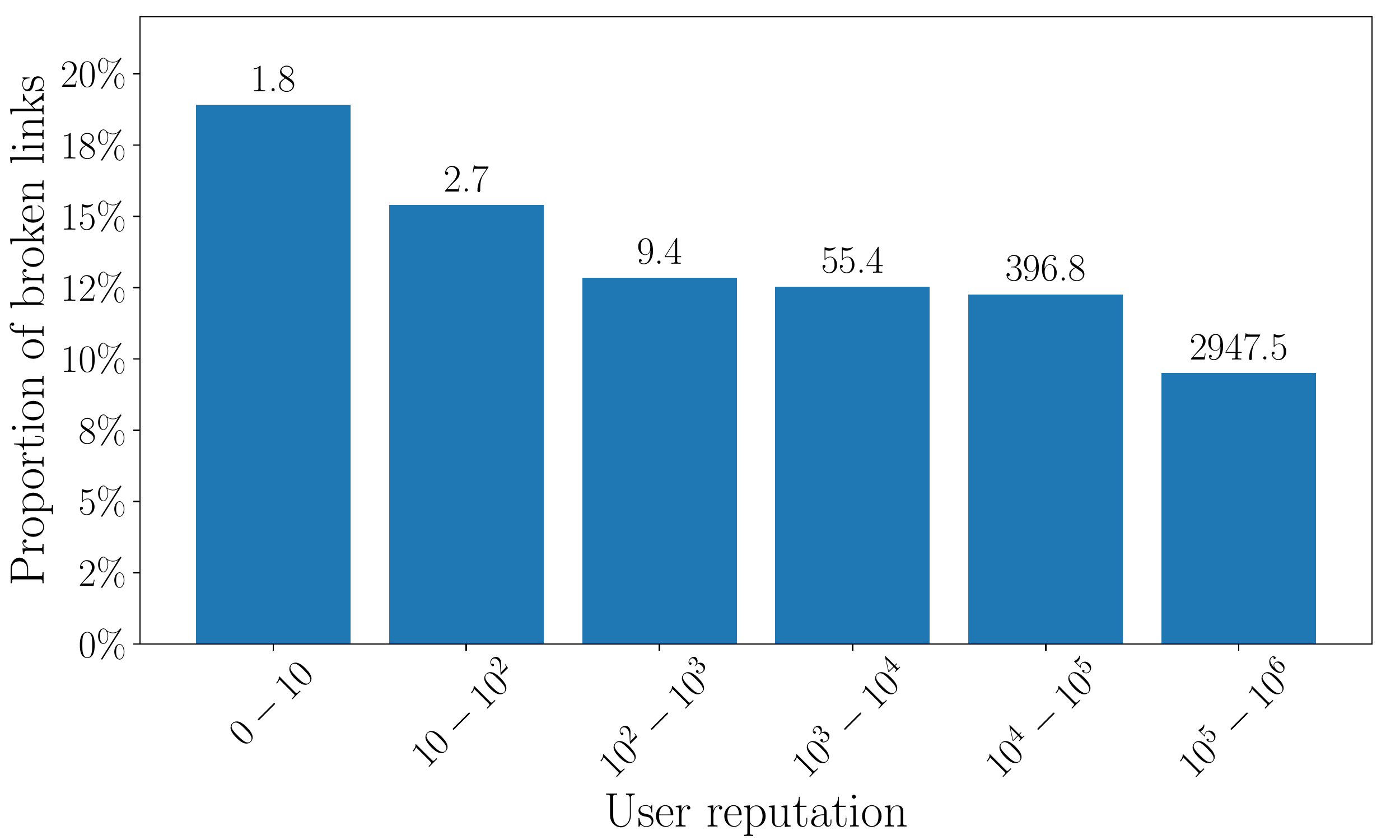}
    \caption{The proportion of broken links among the links posted by the users with different reputations.
    The text label above each bar displays the number of links that are posted by the users with different reputations.
    These figures show that broken links are less common in users with a higher reputation.
    }\label{figure_user_reputation}
\end{figure}

Figure \ref{figure_user_reputation} shows the proportions of broken links among the links posted by the users with different reputations.
The proportions decrease from 18.6\% (for the users with a reputation of less than 10) to 7.4\% (for the users with a reputation higher than $10^5$).
More specifically, we observe that the proportions of broken links among the links posted by different users and the user reputation are significantly correlated with Pearson's correlation coefficient = -0.94 (p-value $<$ 0.05) \cite{benesty2009pearson}.
\textbf{This shows that users with higher reputations are associated with fewer broken links, i.e., the links posted by users with a higher reputation are more likely to be permanent links.}
We suggest viewers use user reputations to detect broken links, as viewers are less likely to encounter broken links in the posts that are posted by the users with higher reputations.
One possible reason is that the users with lower reputations ask only one or two questions with a limited number of links, e.g., the links to show examples.
In Figure \ref{figure_user_reputation}, the text label above each bar displays the number of links that are posted by the users with different reputations.
More specifically, we observe that the proportions of broken links among the links posted by different users and the number of links posted by users with different reputations are significantly correlated with Spearman's rank correlation coefficient = -1.0 (p-value $<$ 0.05) \cite{fieller1957tests}.
One link being broken leads to a large proportion of broken links among the links posted by the users with lower reputations.
In contrast, the users with higher reputations usually answer questions more frequently with a larger number of links.
One link being broken leads to a small proportion of broken links among the links posted by the users with higher reputations.

\rqbox{Only 1.57\% of the posts with broken links are highlighted as such by viewers in the posts' comments.
Only 5.8\% of the posts with broken links removed the broken links.
Viewers vote up more on the posts without broken links than the posts with broken links after browsing.
Users with higher reputation are associated with fewer broken links.
Viewers cannot fully rely on the vote scores to detect broken links, as broken links are common across posts with different vote scores.
}

\subsection{Which Websites are Referenced by Broken Links the Most on Stack Overflow?}\label{findings_websites}

\vspace{0.1cm}\noindent\textbf{Motivation:}
It is still unclear which websites are referenced by broken links on Stack Overflow the most.
By understanding the websites referenced by broken links, Stack Overflow could pay more attention to the websites that are referenced by broken links the most.

\vspace{0.1cm}\noindent\textbf{Approach:}
To understand the websites that are referenced by broken links on Stack Overflow the most, we perform both quantitative and qualitative analysis.
In the quantitative analysis, we captured an overall picture of the websites referenced by broken links.
More specifically, we first quantify the numbers of broken links that reference to different websites and analyze their distribution.
For the websites referenced by different number of links on Stack Overflow, we also analyze whether the websites referenced by more links on Stack Overflow are more likely to be referenced by broken links.

We analyze the websites that are referenced by broken links the most to summarize the types of root causes of the broken links and suggest the corresponding detection and fixing strategies.
We take the top 20 websites ordered by the number of broken links referencing to them as an example.
The number of broken links referencing to the top 20 websites accounts for 27.6\% of the broken links on Stack Overflow.
Table \ref{broken_table} shows the top 20 websites ordered by the number of broken links referencing to them.
We also present the dominant response code.
To analyze the website types of the top 20 websites ordered by the number of broken links referencing to them, we manually performed a lightweight open coding process \cite{seaman_1999}.
This process involves 3 phases and is performed by the first two authors (i.e., A1 and A2) of this paper:

\begin{itemize}[wide=0pt]
    \item Phase I: The first two authors independently categorized the types of websites. 
    For the websites that can be accessed currently, the first two authors checked the content of the websites.
    For the websites that cannot be visited currently, the first two authors referred to related descriptions.
    For example, the first two authors searched the website on Google and read the web pages that are presented in search results, e.g., the Internet Archive's Wayback Machine\footnote{https://archive.org/web/}.
    The first two authors took notes regarding the deficiency or ambiguity of the labeling for these websites.
    Table \ref{table_website_type} presents the types of websites identified in this phase.
    \item Phase II: The first two authors discussed the coding results to resolve any disagreements until a consensus was reached.
    For example, A1 considered \lstinline{social.msdn.microsoft.com} as a documentation website because this website is the subdomain of the Microsoft documentation website (i.e., \lstinline{msdn.microsoft.com}).
    A2 considered \lstinline{social.msdn.microsoft.com} as a forum website because 90\% of the links to this website connect to the \lstinline{https://social.msdn.microsoft.com/Forums} sub-path.
    Finally, we consider \lstinline{social.msdn.microsoft.com} as a forum website, and the resources hosted in this website can be maintained by the users and moderators.
    The interrater agreement of this coding process has a Cohen's kappa of 0.93 (measured before discussion), indicating that the agreement level is high \cite{viera2005understanding}.
\end{itemize}

\begin{table*}[htb] 
    \centering
    \footnotesize
    \caption{The website types of the top 20 websites in terms of the number of broken links referencing to them.
    }\label{table_website_type}
\begin{tabular}{|l|l|l|l|}
\hline
    \textbf{Type}  & \textbf{Function}                                     & \textbf{Maintainer}   & \textbf{Example}   \\ \hline
    Code                   & Share code projects, code snippets, and runnable code examples.   & users  & github.com, pastebin.com, jsfiddle.net \\ \hline
    Documentation          & Share official development related documentation of a product.     & official teams & docs.oracle.com                      \\ \hline
    Official               & Share a starting point to other resources of the product.         & official teams  & www.microsoft.com            \\ \hline
    File Hosting           & Provide file hosting services.                                & users     & www.dropbox.com                      \\ \hline
    Image Hosting          & Provide online image sharing and hosting services.           & users      & i.stack.imgur.com                \\ \hline
    \end{tabular}
\end{table*}

\begin{figure}[!htb]
    \centering
    \includegraphics[width = 0.5\textwidth]{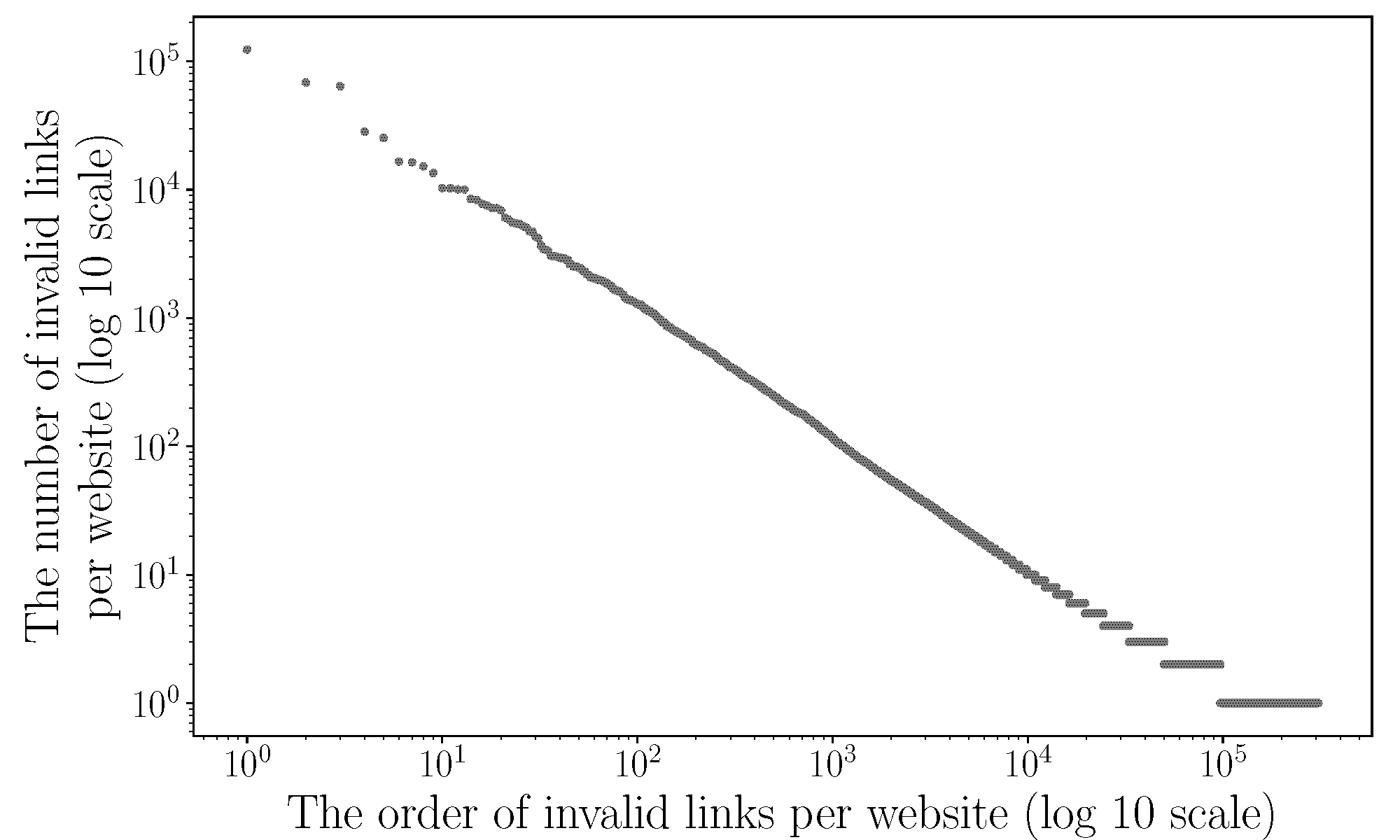}
    \caption{The distribution of the numbers of broken links that reference to different websites in descending order.
    These figures show that the numbers of the broken links in different websites conform to the power-law distribution.
    }\label{figure_domainpowerlaw}
\end{figure}

\vspace{0.1cm}\noindent\textbf{Results:}
\subsubsection{Quantitative Results}
\textbf{50\% (i.e., 844,002) of broken links reference to the top 0.3\% (i.e., 414) websites in terms of the number of the broken links referencing to them.}
308,737 (i.e., 46.9\%) of the websites that are referenced by Stack Overflow are referenced by broken links.
Figure \ref{figure_domainpowerlaw} shows the plot of the numbers of broken links that reference to different websites on Stack Overflow.
The numbers of the broken links that reference to different websites conform to the power-law distribution with $\alpha=1.96$ and $xmin=10.0$.

\begin{figure}[!htb]
    \centering
    \includegraphics[width = 0.5\textwidth]{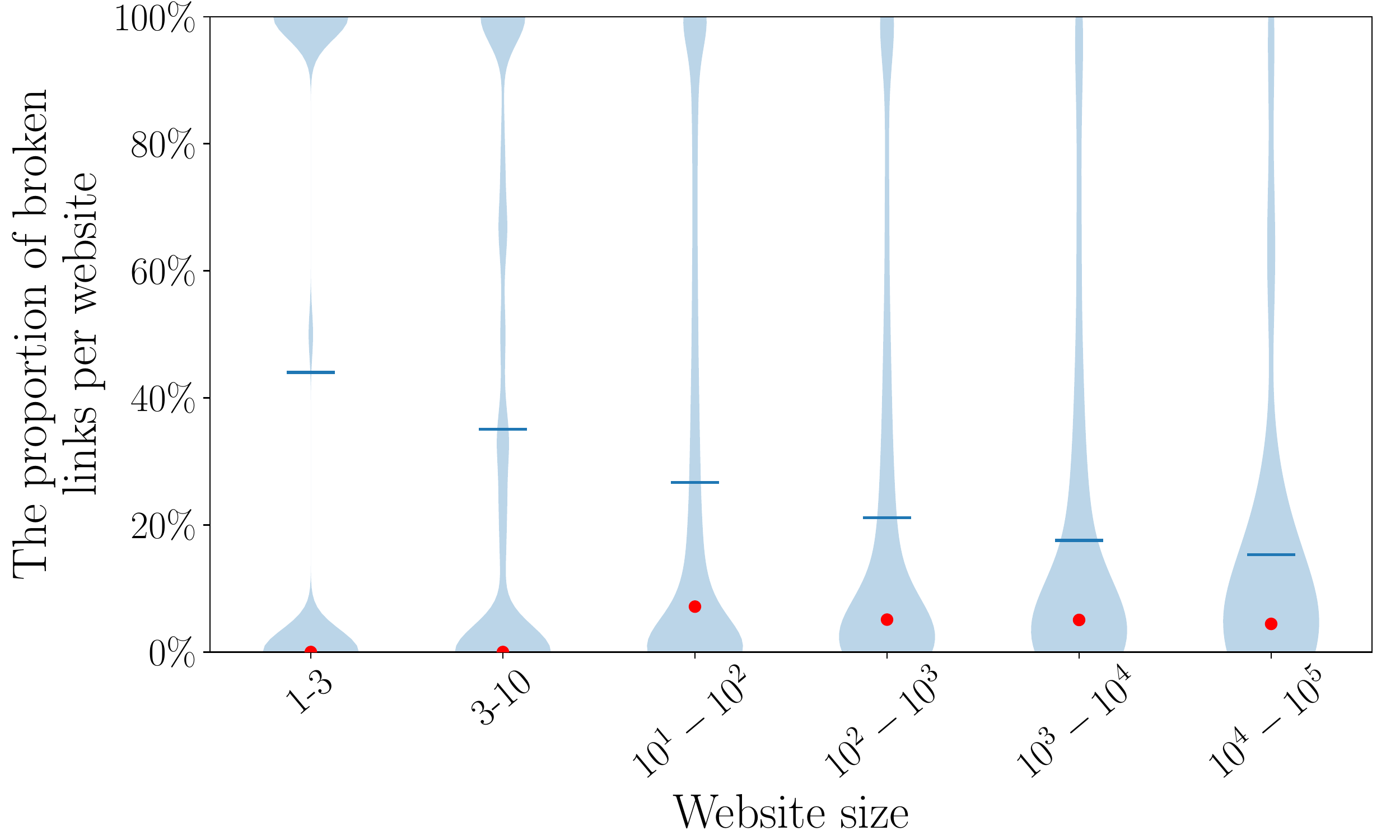}
    \caption{The distribution of the proportions of broken links among the links that reference to different websites based on the number of the links that reference to the website on Stack Overflow.
    This figure shows that websites that are referenced by fewer links on Stack Overflow are more likely to have broken links.
    }\label{figure_domainsizeprop}
\end{figure}

Figure \ref{figure_domainsizeprop} plots the proportions of broken links among the links that reference to different websites based on the number of the links that reference to the website on Stack Overflow.
To check whether the differences in the proportions of broken links among the links that reference to different websites are statistically significant between the websites with different number of links that reference to them, we perform the Kruskal-Wallis H-test \cite{kruskal1952use}.
The null hypothesis is that there is no difference between the websites with different numbers of links that reference to them in terms of the proportion of broken links.
As a result, the differences between the websites with different numbers of links that reference to them are significant (p-value $<$ 0.05).
We then calculate Cliff's delta to measure the effect size \cite{cliff2014ordinal}.
As a result, the differences between the websites with different numbers of links that reference to them is small in terms of the proportions of broken links among the links that reference to them (Cliff's delta is between 0.147 and 0.33).
More specifically, we observe that \textbf{websites that are referenced by fewer links on Stack Overflow are more likely to be referenced by broken links.}
For example, \lstinline{github.com} is the second most commonly shared external website on Stack Overflow (1,870,707 links on Stack Overflow reference to \lstinline{github.com}).
6.6\% of the links that reference to \lstinline{github.com} are broken links.
The broken links that reference to \lstinline{github.com} account for 7.3\% of the broken links on Stack Overflow.
In contrast, among 468,577 links that reference the websites with 1--3 links that are shared on Stack Overflow, 43\% of them are broken links.
This proportion is 6.5 times higher than the proportion of broken links among the links that reference to \lstinline{github.com}.
The broken links that reference the websites with 1--3 links that are shared on Stack Overflow account for 16\% of the total number of broken links.
This indicates that the websites with 1--3 links that are shared on Stack Overflow contribute 2.2 times broken links on Stack Overflow more than \lstinline{github.com}.
The websites that are referenced by more links on Stack Overflow are less likely to be referenced by broken links in terms of their proportion.

\begin{figure}[!htb]
    \centering
    \includegraphics[width = 0.5\textwidth]{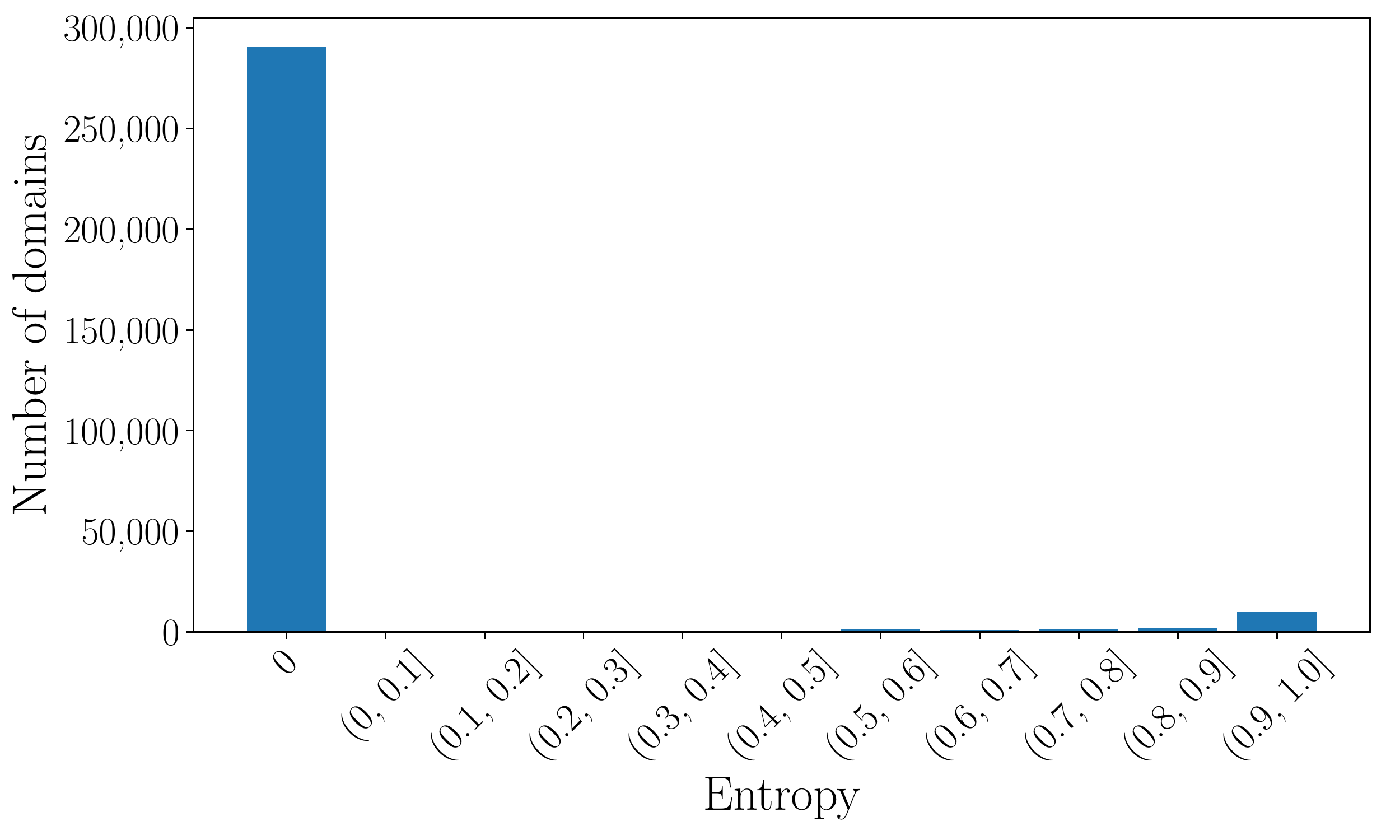}
    \caption{The distribution of the normalized entropy of the response code of the broken links that reference to each website.
    This figure shows that most of the broken links that reference to each website are caused by the same reason.
    }\label{figure_domainentropy}
\end{figure}

Table \ref{broken_table} shows the dominant response code of the broken links that reference to different websites on Stack Overflow.
Over 90\% of the broken links that reference to 12 websites are responded with the same response code.
For example, 93.15\% of the broken links that reference to jsfiddle.net are responded with 404 response code.
To test whether most of the broken links that reference to the same websites are responded with the same response code is common in Stack Overflow, for each website that are referenced by broken links on Stack Overflow, we calculate its the normalized entropy of the response code of the broken links.
Different from a simple statistic of the total number of hits by error codes suffice, normalized entropy can describe the randomness of the information in the groups with different sizes.
A non-uniform distribution would have less normalized entropy (i.e., less random) than a uniform distribution.
In this paper, we use the normalized entropy to measure the randomness of the response code of the broken links in each website.
We calculate the normalized entropy by dividing the entropy with the number of broken links that point to that website.
Figure \ref{figure_domainentropy} plots the normalized entropy of the response code of the broken links that reference to each website.
As a result, most of the normalized entropies are 0, indicating that \textbf{most of the broken links that reference to different website are caused by the same reason}.

\subsubsection{Qualitative Results}

\begin{table*}[!htb]
    \centering
    \caption{The top 20 websites in terms of the number of broken links on Stack Overflow.
    }\label{broken_table}
    \begin{tabular}{|l|l|p{0.15\textwidth}|p{0.15\textwidth}|p{0.15\textwidth}|p{0.12\textwidth}|}
    \hline
    \textbf{Website}          & \textbf{Website Type} & \textbf{\% among all broken links} & \textbf{\% in website} & \textbf{Dominant Status Code} & \textbf{\% Dominant Status Code} \\ \hline
    github.com                & code                  & 7.33\%                              & 6.62\%                 & 404                           & 82.97\%                          \\ \hline
    codepen.io                & code                  & 4.05\%                              & 63.68\%                & 403                           & 97.80\%                          \\ \hline
    pastebin.com              & code                  & 3.81\%                              & 58.32\%                & 403                           & 95.96\%                          \\ \hline
    jsfiddle.net              & code                  & 1.50\%                              & 2.13\%                 & 404                           & 93.15\%                          \\ \hline
    code.google.com           & code                  & 0.97\%                              & 8.67\%                 & 405                           & 76.72\%                          \\ \hline
developer.apple.com       & documentation         & 0.80\%                              & 6.10\%                 & 404                           & 93.17\%                          \\ \hline
    gist.github.com           & code                  & 0.61\%                              & 9.75\%                 & 404                           & 80.84\%                          \\ \hline
    grepcode.com              & code                  & 0.60\%                              & 83.11\%                & TCPTimedOutError              & 62.06\%                          \\ \hline
    msdn.microsoft.com        & documentation         & 0.60\%                              & 1.08\%                 & 503                           & 49.75\%                          \\ \hline
    social.msdn.microsoft.com & forum         & 0.50\%                              & 22.75\%                & Error                         & 98.16\%                          \\ \hline
    www.microsoft.com         & official              & 0.49\%                              & 21.65\%                & 404                           & 61.49\%                          \\ \hline
    pastie.org                & code                  & 0.46\%                              & 91.14\%                & 500                           & 47.39\%                          \\ \hline
    dl.dropboxusercontent.com & file hosting          & 0.45\%                              & 90.63\%                & 404                           & 97.84\%                          \\ \hline
    dl.dropbox.com            & file hosting          & 0.43\%                              & 90.48\%                & 404                           & 98.94\%                          \\ \hline
    postimg.org               & image sharing         & 0.41\%                              & 94.28\%                & DNSLookupError                & 100.00\%                         \\ \hline
    docs.oracle.com           & documentation         & 0.34\%                              & 1.49\%                 & 404                           & 94.03\%                          \\ \hline
    docs.djangoproject.com    & documentation         & 0.33\%                              & 6.53\%                 & 404                           & 99.98\%                          \\ \hline
    drive.google.com          & file hosting          & 0.32\%                              & 22.07\%                & 404                           & 96.38\%                          \\ \hline
    fiddle.jshell.net         & code                  & 0.32\%                              & 73.38\%                & 404                           & 99.78\%                          \\ \hline
    ideone.com                & code                  & 0.30\%                              & 7.91\%                 & 404                           & 98.54\%                          \\ \hline
\end{tabular}
\end{table*}

Table \ref{broken_table} shows the top 20 websites ordered by the number of broken links referencing to them, as well as the dominant response code.
\textbf{Among the top 20 websites ordered by the number of broken links referencing to them, 15 websites host the resources that can be maintained by their users.}
For example, resources hosted in 10 \textit{code} websites, 3 \textit{file-hosting} websites, 1 \textit{forum} websites, and 1 \textit{image} websites can be maintained by their users, such as deleting the resources that can be referenced by links.
8 of the websites that host the resources that can be maintained by their users are mainly responded with a 404 response code, i.e., a client-side error and indicating that the resources hosted on that links cannot be found.
This shows that one of the possible reasons for the broken links that reference to the websites that can be maintained by their users is that users can delete the resources according to their judgment.
For example, the comments to a post\footnote{https://stackoverflow.com/q/9065853/} that answers how to auto-resize the input field with jQuery indicate that

\vspace{0.1cm} 
\hangindent 1.5em\textit{That link leads to GitHub's 404 page.}  – CoderDennis \newline
\hangindent 1.5em\textit{Yeah, the author removed it. See my updated answer.}  – Dmitry Pashkevich 
\vspace{0.1cm}
  
\noindent\normalsize
This comment shows that the resources that are hosted on other websites (i.e., GitHub) can be removed by the owner of the resources.
To help viewers avoid broken links when browsing Stack Overflow, we suggest viewers be cautious about the links that host resources that can be maintained by their users.
To maintain the crowdsourced knowledge on Stack Overflow, we suggest that Stack Overflow should archive snapshots of links to backup the resources that are maintained by users.

\textbf{Posts with links that reference to code websites are the ones that are the least maintained.}
For example, \textit{github.com} are referenced by the largest number of broken links on Stack Overflow.
123,774 (i.e., 6.6\%) of all broken links on Stack Overflow reference to \textit{github.com}.
The resources hosted on \textit{github.com} is maintained by their users.
82.97\% of the broken links reference to this website are responded with 404, indicating that these resources are removed from \textit{github.com}.
We suggest the Stack Overflow users not reference to the resources hosted on code websites.
We encourage users to paste code within the Stack Overflow websites, e.g., using code blocks or Stack Snippets \cite{code_editing, stack_snippets}.

Resources hosted in 4 \textit{documentation} websites and 1 \textit{official} website are maintained by the official teams.
For example, \textit{developer.apple.com} is referenced by broken links the most among the websites that are maintained by the official teams.
This website is the least maintained documentation website among all the websites referenced on Stack Overflow.
One possible reason for the broken links that reference to the developer.apple.com website is related to the deprecation of APIs.
For example, a comment to an accepted answer\footnote{https://stackoverflow.com/q/1252397/} to where to get the standalone executable binary CVS for OSX indicates that

\vspace{0.1cm} 
\hangindent 1.5em\textit{Sadly, as of Feb 2014, those are both dead links. According to \textit{developer.apple.com/library/ios/releasenotes/DeveloperTools/…}\footnote{https://developer.apple.com/library/ios/releasenotes/Developer-Tools/RN-Xcode/} CVS and RCS have been removed as of Xcode 5. The new official CVS web page seems to be \textit{savannah.nongnu.org/projects/cvs }\footnote{http://savannah.nongnu.org/projects/cvs}}
\vspace{0.1cm}
  
\noindent\normalsize
This comment shows that the webpages that host the deprecated APIs are directly removed without being archived. 
We suggest the designers of the websites that are maintained by the official teams could redirect the requests for the outdated web pages to an updated web page.

\rqbox{
    50\% (i.e., 844,002) of broken links reference to the top 0.3\% (i.e., 414) websites in terms of the number of the broken links referencing to them.
    Websites that are referenced by fewer links on Stack Overflow are more likely to be referenced by broken links.
    The websites that host resources maintained by their users are the least maintained, e.g., github.com.
}

\subsection{Are the posts and comments Associated with Particular Tags More Likely to Have Broken Links than Others?}\label{findings_tags}

\vspace{0.1cm}\noindent\textbf{Motivation:}
It is still unclear the posts and comments associated with which tags suffer from the broken links the most on Stack Overflow.
By understanding the severity of the broken links problem associated with different Stack Overflow tags, we could suggest that the viewers who seek solutions of the questions related to certain tags should be more cautious.

\vspace{0.1cm}\noindent\textbf{Approach:}
To find out which tags are the most associated with broken links, we first obtain the question threads (i.e., a Stack Overflow questions, together with its answers and comments) that reference to the broken links.
Then we extract the tags of each question from the Stack Overflow data dump \texttt{Posts}.
In this paper, given an answer or a comment, we use the tags of its corresponding question as its tags.
Finally, we group the broken links into different tags and count the numbers.

\begin{figure}[!htb]
    \centering
    \includegraphics[width = 0.5\textwidth]{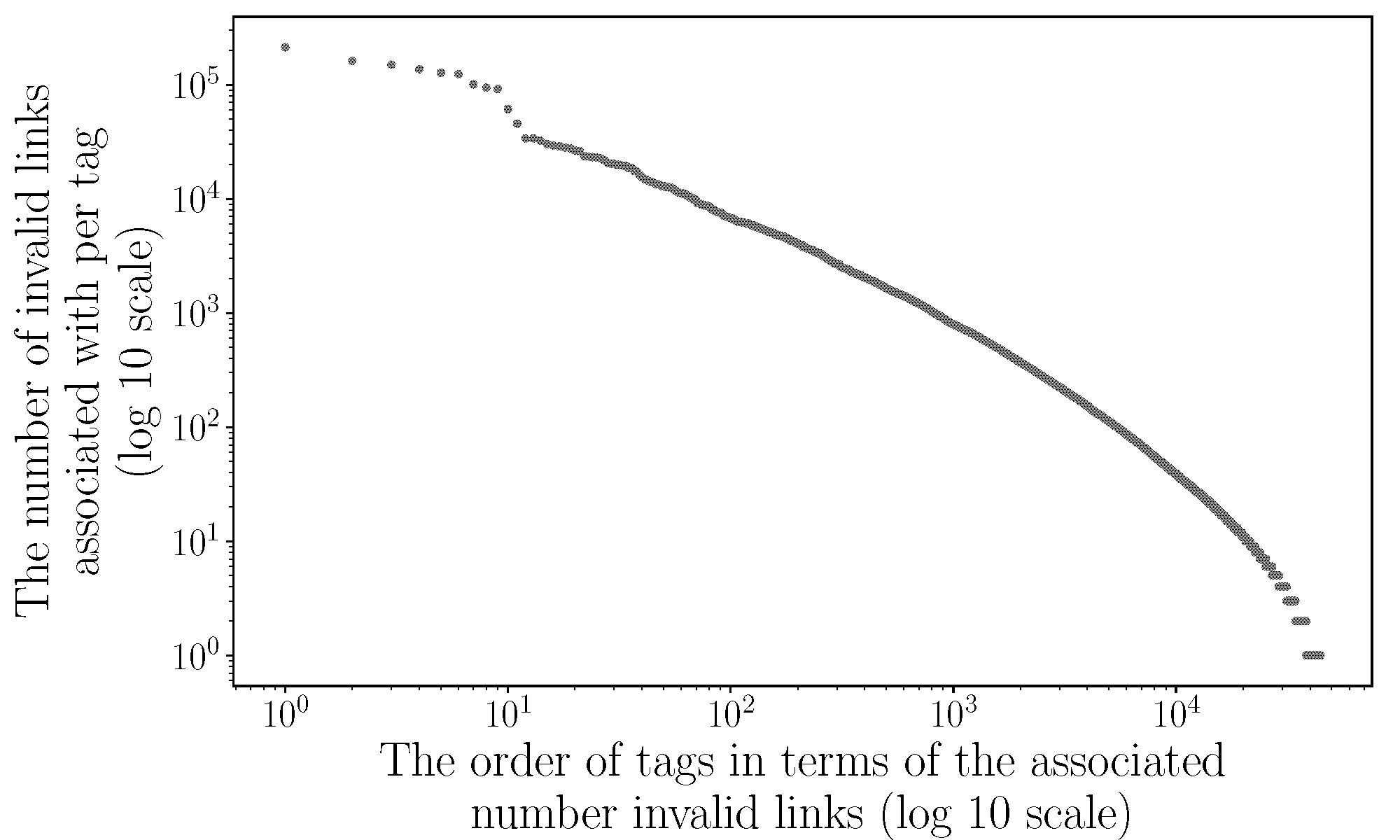}
    \caption{The number of broken links associated with different tags in descending order.
    }\label{figure_tag}
\end{figure}

\vspace{0.1cm}\noindent\textbf{Results:}
Among 55,027 tags on Stack Overflow, the posts and comments related to 54,083 tags contain links, and the posts and comments related to 44,413 (i.e., 82.1\%) tags contain broken links.
Figure \ref{figure_tag} shows the plot of the numbers of broken links associated with different tags.
\textbf{The top 10 tags in terms of the number of associated broken links corresponds to 55.4\% of the broken links on Stack Overflow.}
We suggest Stack Overflow be cautious of the question threads that are associated with the top 10 tags in terms of the number of associated broken links.

\begin{table}[]
    \centering
    \caption{Top 10 tags in terms of the number of broken links.
    This Table shows that web development related technologies are more likely to have broken links.
    }\label{top_10_tag_prop}
    \begin{tabular}{|l|r|r|r|}
    \hline
    \textbf{Tag} & \textbf{\# Broken Links} & \textbf{\# Links} & \textbf{\% Broken Links} \\ \hline
    javascript              & 213,532                   & 1,668,150         & 12.80\%                   \\ \hline
    php                     & 162,454                   & 769,853           & 21.10\%                   \\ \hline
    java                    & 149,973                   & 982,938           & 15.26\%                   \\ \hline
    html                    & 137,063                   & 1,045,135         & 13.11\%                   \\ \hline
    css                     & 127,719                   & 954,516           & 13.38\%                   \\ \hline
    jquery                  & 123,606                   & 970,833           & 12.73\%                   \\ \hline
    c\#                     & 100,880                   & 890,799           & 11.32\%                   \\ \hline
    android                 & 94,348                    & 690,778           & 13.66\%                   \\ \hline
    python                  & 92,122                    & 773,885           & 11.90\%                   \\ \hline
    c++                     & 61,316                    & 472,542           & 12.98\%                   \\ \hline
\end{tabular}
\end{table}

Table \ref{top_10_tag_prop} shows the top 10 tags in terms of the number of the associated broken links.
\textbf{The posts and comments related to the web technologies, i.e., JavaScript, HTML, CSS, and jQuery, are associated with more broken links.}
When Stack Overflow viewers browse the posts and comments related to the knowledge on web technologies, they are more likely to cannot fully understand the questions.
We suggest that Stack Overflow should pay more attention to the posts and comments related to web technologies.

\begin{figure}[!htb]
    \centering
    \includegraphics[width = \linewidth]{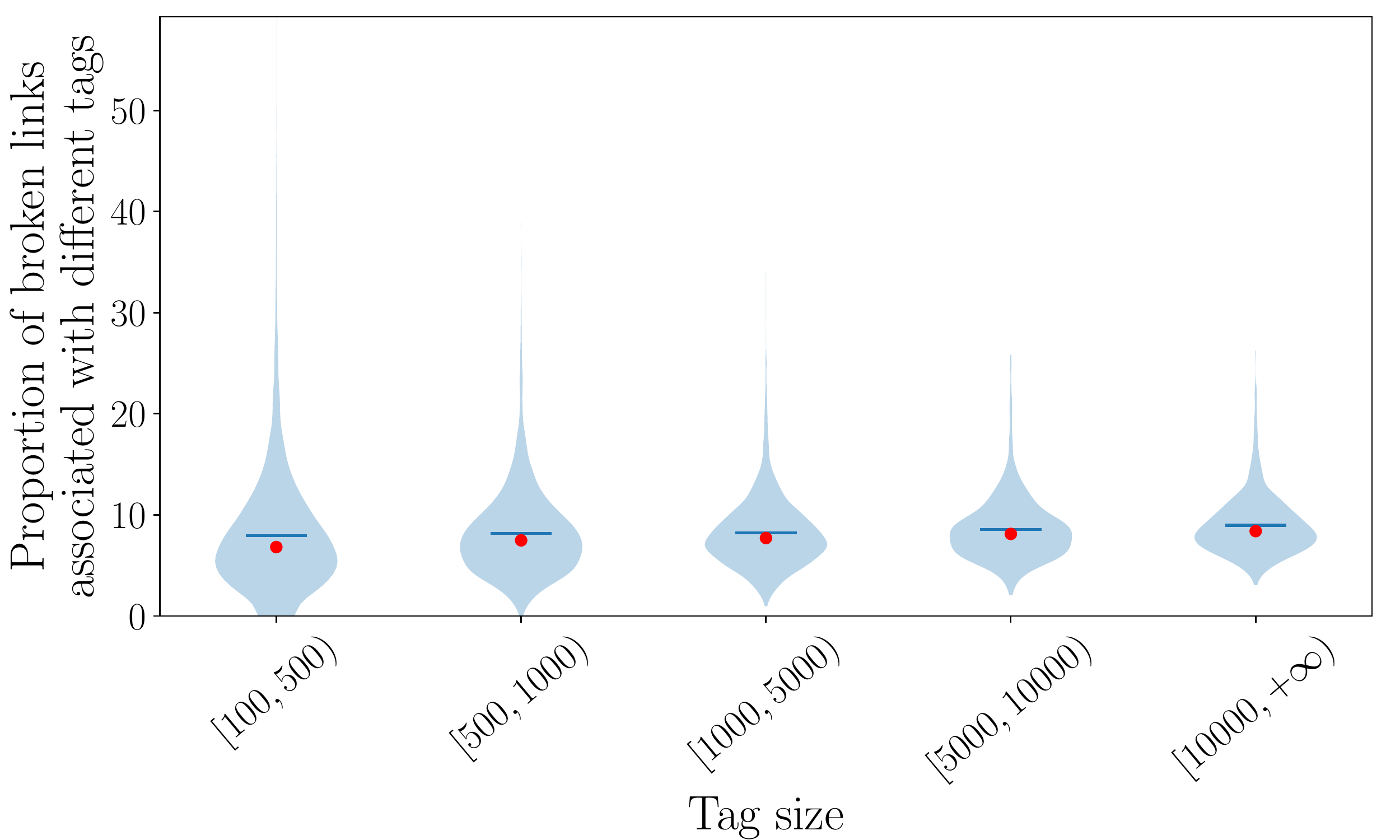}
    \caption{The proportions of broken links among the links that are associated with the tags with different numbers of question threads.
    This figure shows that broken links are more common for popular tags on Stack Overflow.
    }\label{figure_tagsize}
\end{figure}

Figure \ref{figure_tagsize} shows the violin plots of the proportions of broken links among the links that are associated with the tags with the different numbers of question threads.
To check whether the differences in the proportions of broken links among the links that are associated with different tags are statistically significant between the tags with the different numbers of question threads, we perform the Kruskal-Wallis H-test \cite{kruskal1952use}.
The null hypothesis is that there is no difference between the tags with the different numbers of question threads in terms of the proportions of broken links.
As a result, the differences between the tags with the different numbers of question threads are significant (p-value $<$ 0.05). We then calculate Cliff's delta to measure the effect size \cite{cliff2014ordinal}.
As a result, the differences between the tags with the different numbers of question threads are small in terms of the proportions of broken links (Cliff's delta is between 0.147 and 0.33).
More specifically, we observe that the tags with 100 to 500 questions threads have the least proportion of broken links, and the tags with over 10,000 questions threads have the largest proportion of broken links.
This shows that \textbf{broken links are more common for popular tags on Stack Overflow.}

\begin{figure}[!htb]
    \centering
    \includegraphics[width = \linewidth]{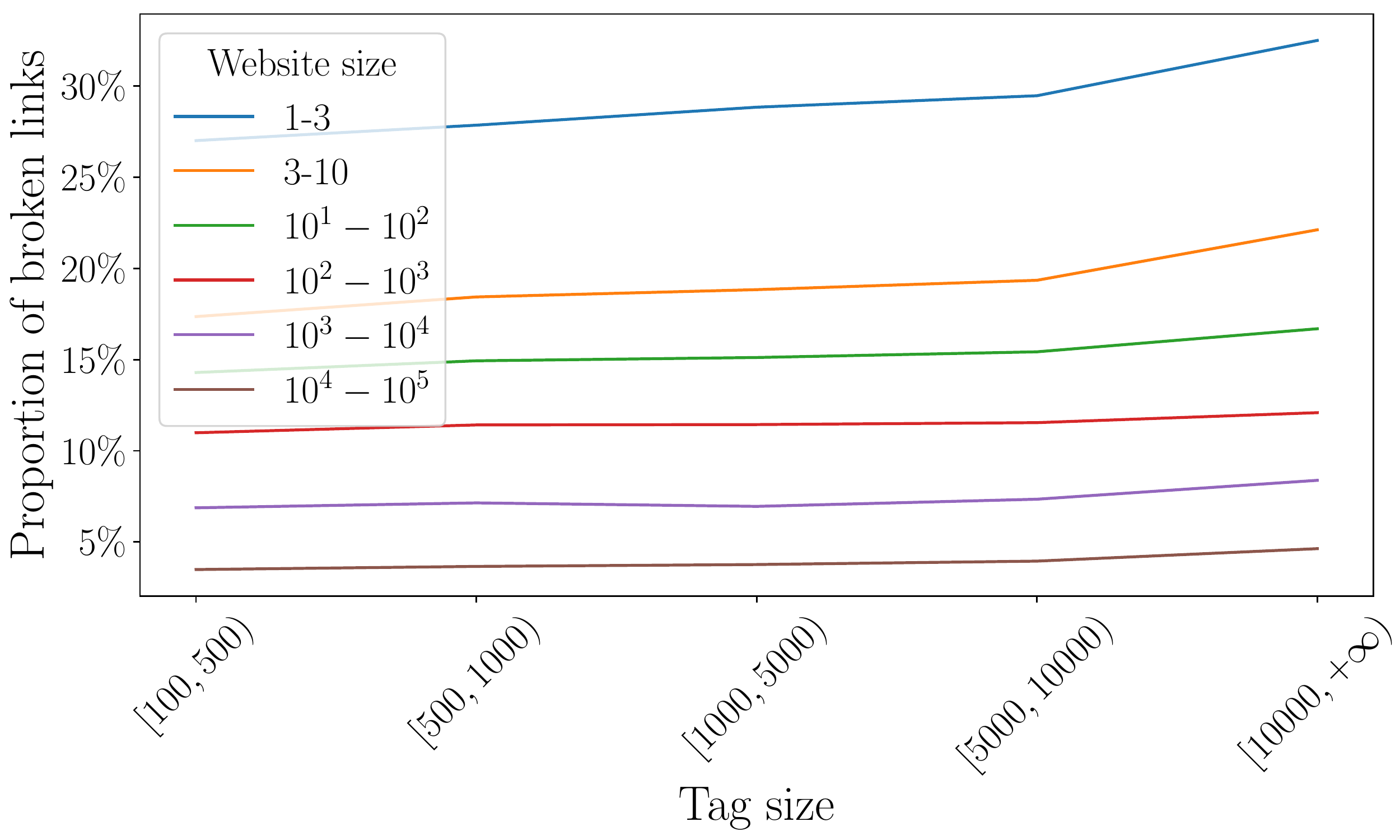}
    \caption{
        For the tags with different sizes (i.e., the number of question threads that are associated with certain tags), this figure plots the proportion of broken links among the links that reference to different sizes of websites (i.e., the number of links that reference to the specific website on Stack Overflow). 
        This figure shows that the links that reference to smaller websites on Stack Overflow are more likely to be broken in the popular tags.
    }\label{figure_tag_domain_size}
\end{figure}

For the tags with different sizes (i.e., the number of question threads that are associated with certain tags), Figure \ref{figure_tag_domain_size} plots the proportion of broken links among the links that reference to different sizes of websites (i.e., the number of links reference to the website on Stack Overflow).
We observe that the average proportions of broken links among the links that reference to the websites with 1--3, 3--10, $10^1$--$10^2$, $10^2$--$10^3$, $10^3$--$10^4$, $10^4$--$10^5$ links that are shared on Stack Overflow and the sizes of tags are significantly correlated with Pearson's correlation coefficients from 0.92 (for $10^4$--$10^5$) to 0.98 (for $10^1$--$10^2$) (p-values $<$ 0.05) \cite{benesty2009pearson}.
The average proportions of broken links among the links that reference to the websites with $10^6$--$10^7$ links that are shared on Stack Overflow and the sizes of tags are \textbf{not} significantly correlated with Pearson's correlation coefficient = 0.81 (p-value = 0.10) \cite{benesty2009pearson}.
For example, for the tags with 100–500 question threads, the proportion of broken links among the links that reference to the websites with 1–3 links that are shared on Stack Overflow is 39.4\%.
But for the tags with over 10,000 question threads, the proportion of broken links among the links that reference to the websites with 1–3 links that are shared on Stack Overflow is 46.4\% (i.e., 1.18 times higher than that in the tags with 100–500 question threads).
For the tags with 100–500 question threads, the proportion of broken links among the links that reference to the websites with over 10,000 links that are shared on Stack Overflow is 6.5\%.
But for the tags with over 10,000 question threads, the proportion of broken links among the links that reference to the websites with over 10,000 links that are shared on Stack Overflow is 6.9\% (i.e., 1.07 times higher than that in the tags with 100–500 question threads).
This shows that \textbf{in popular tags, the broken links are more likely to reference to the websites that are referenced by fewer links on Stack Overflow.}

\begin{figure}[!htb]
    \centering
    \includegraphics[width = \linewidth]{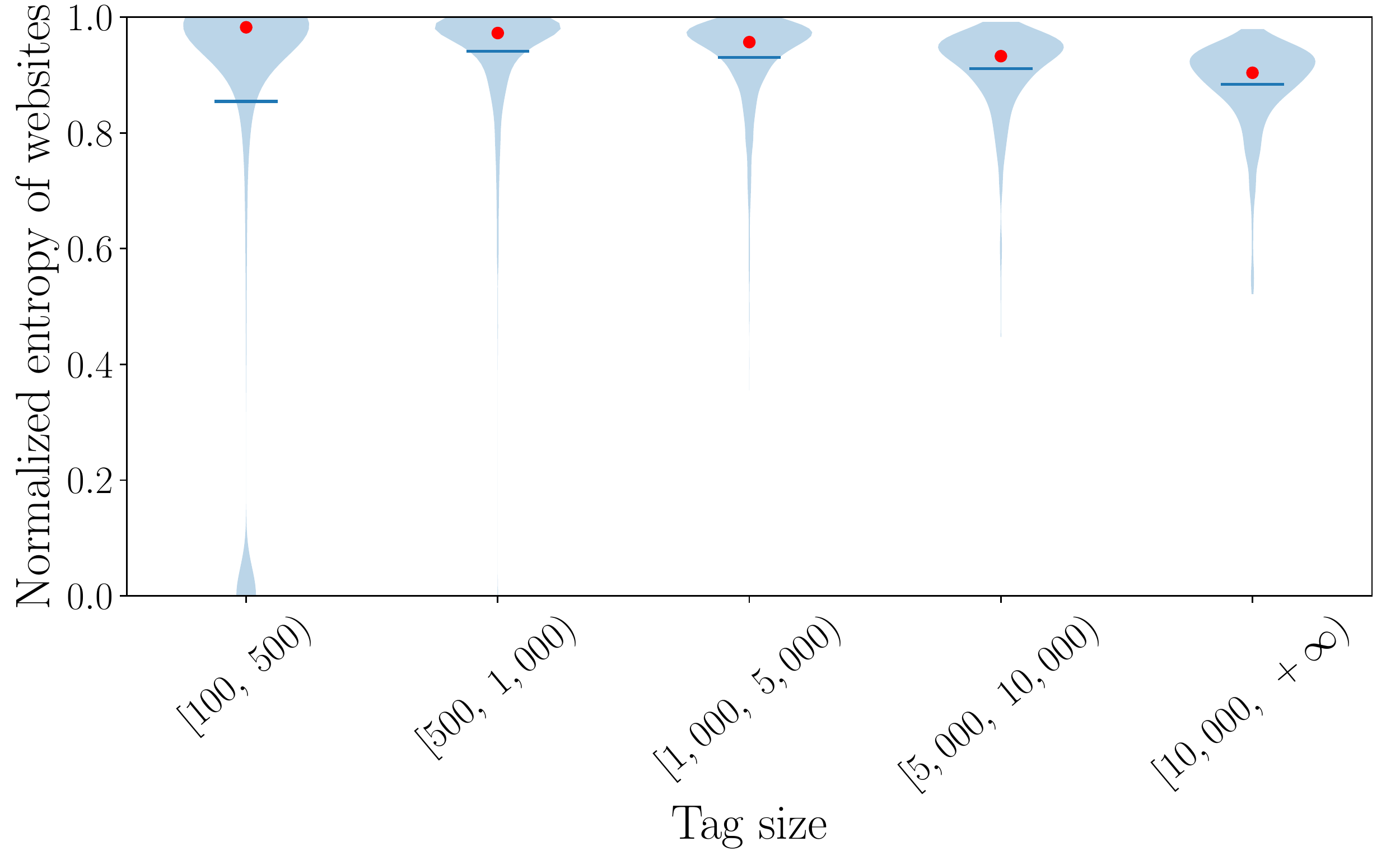}
    \caption{
        The distribution of the normalized entropy of the websites that are associated with the broken links based on the tags with different sizes.
        This figure shows that the broken links associated with most tags are uniformly referencing to different websites.
        The broken links associated with popular tags are less likely to uniformly referencing to different websites.
    }\label{figure_tag_domain_entropy}
\end{figure}

Figure \ref{figure_tag_domain_entropy} shows the distribution of the normalized entropy of the websites that are associated with each tag.
More specifically, the normalized entropies of 75\% tags are higher than 0.8 in terms of the websites that are associated with them.
This indicates that \textbf{broken links associated with most tags are uniformly referencing to different websites.}
This shows that focusing on repairing the broken links that reference to specific websites cannot help with the broken links problem associated with a certain tag.
To check whether the differences between the tags with different numbers of question threads are significant in the normalized entropies of the websites that are associated with the specific tag, we perform the Kruskal-Wallis H-test \cite{kruskal1952use}.
The null hypothesis is that there is no difference between the tags with different numbers of question threads in terms of the entropies of the websites that are associated with them.
As a result, the differences between the tags with different numbers of question threads in terms of the normalized entropies of the websites that are associated with them are significant (p-value < 0.05).
We then compute Cliff's delta to measure the effect-size \cite{cliff2014ordinal}.
As a result, the differences between the tags with different numbers of question threads in terms of the normalized entropies of the websites that are associated with them are large (Cliff's delta > 0.474).
This shows that the \textbf{broken links in popular tags are less likely to uniformly referencing to different websites}, i.e., the broken links in popular tags are more likely to centrally referencing a limited number of websites.”

\begin{figure}[!htb]
    \centering
    \includegraphics[width = \linewidth]{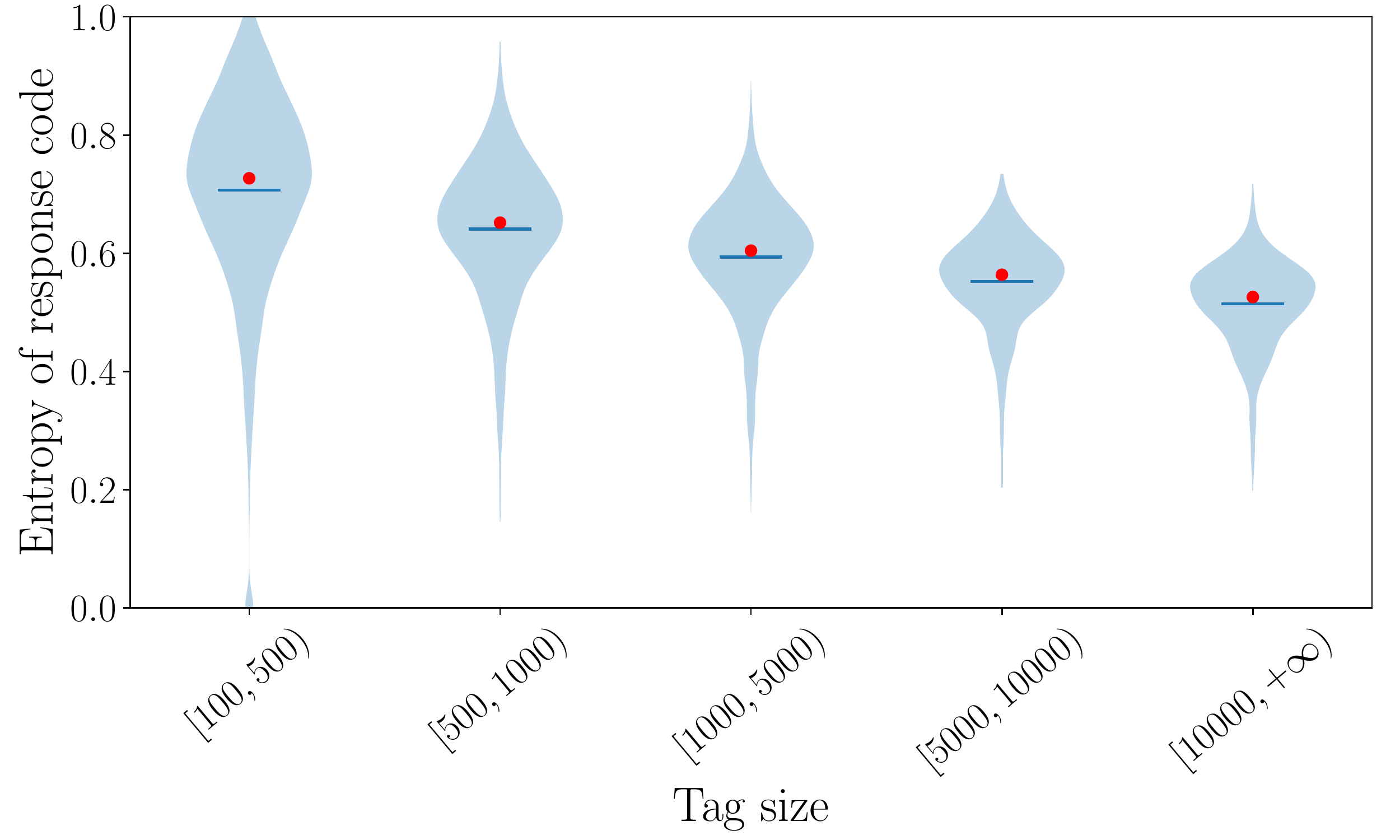}
    \caption{
        The distribution of the normalized entropies of the response code of the broken links that are associated with the tags with different numbers of question threads.
        This figure shows that the response code of the broken links that are associated with the tags with more question threads is less random.
    }\label{figure_tag_responsecode_entropy}
\end{figure}

Figure \ref{figure_tag_responsecode_entropy} shows the distribution of the normalized entropy of the response code of the broken links that are associated with the tag with different numbers of question threads.
To check whether the differences between the tags with different numbers of question threads are significant in the normalized entropies of the response code of the broken links that are associated with the specific tag, we perform the Kruskal-Wallis H-test \cite{kruskal1952use}.
The null hypothesis is that there is no difference between the tags with different numbers of question threads in terms of the entropies of the response code of the broken links that are associated with them. As a result, the differences are significant (p-value < 0.05).
We then compute Cliff's delta to measure the effect-size \cite{cliff1993dominance}.
As a result, the differences between the tags with different numbers of question threads in terms of the normalized entropies of the response code of the broken links that are associated with them are large (Cliff's delta > 0.474).
\textbf{The broken links in popular tags are less likely to be associated with different response codes.}
One possible reason is that the broken links in popular tags are more likely to centrally referencing a limited number of websites and most of the response code of the broken links that reference to the same websites are the same.

\begin{figure}[!htb]
    \centering
    \includegraphics[width = \linewidth]{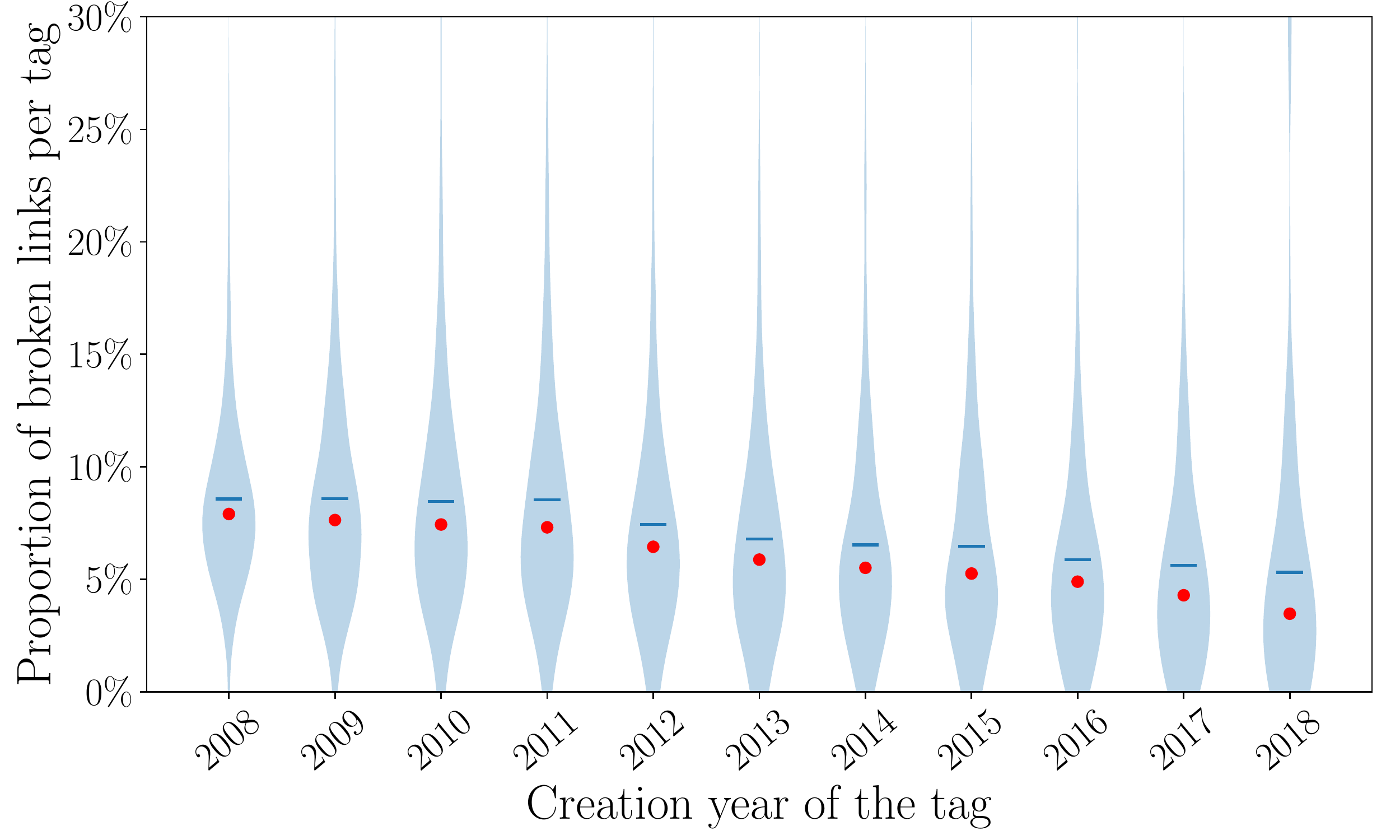}
    \caption{The proportions of broken links among the links that are associated with the tags that were created in different years (we identify the creation of a tag as the first question that is marked with the tag were posted on Stack Overflow).
    This figure shows that older tags are more likely to have broken links.
    }\label{figure_tagdate}
\end{figure}

Figure \ref{figure_tagdate} shows the violin plots of the proportions of broken links among the links associated with different tags based on the creation time of tags (i.e., when the first question that is marked with the tag were posted on Stack Overflow).
To check whether the differences in the proportions of broken links among the links associated with different tags are statistically significant between the tags with different creation dates, we perform the Kruskal-Wallis H-test \cite{kruskal1952use}.
The null hypothesis is that there is no difference between tags with different creation dates in terms of the proportions of broken links.
As a result, the differences between the tags with different creation dates are significant (p-value $<$ 0.05).
We then compute Cliff's delta to measure the effect-size \cite{cliff1993dominance}.
As a result, the difference between tags with different creation dates in terms of the proportions of broken links are large (Cliff's delta $>$ 0.474).
More specifically, \textbf{older tags have a larger proportion of broken links compared with the younger tags.}
One possible reason is that the question threads associated with older tags still host the knowledge that is posted to solve the problems a long time ago.
These nuggets of knowledge may be out-of-date, and the websites that host the out-of-date knowledge may not be maintained anymore.

\rqbox{
    50\% of the broken links are referenced in the threads that are assigned at least one of the following 10 tags: JavaScript, PHP, Java, HTML, CSS, jQuery, C\#, Android, Python, and C++.
    Posts related to web technologies, e.g., JavaScript, HTML, CSS, and jQuery, are associated with more broken links.
    We also observe that popular tags and older tags are more likely to have broken links.
}

\section{Discussion}\label{dis}
In this section, we discuss the implications of our findings for Stack Overflow moderators, users, and researchers. We also consider the threats to the validity of our results.

\subsection{Implications}

\subsubsection{Actionable Suggestions for Researchers}\label{imp_res}

\textbf{Future work could repair the broken links on Stack Overflow based on the revisions of links.}
In Section \ref{findings_mechanism}, we observe that broken links have negative impacts on the crowdsourced knowledge on Stack Overflow.
We also observe that 2,458,323 revisions repaired the broken links.
For example, the broken link \lstinline{http://dev.mysql.com/doc/refman/5.5/en/reserved-words.html} is replaced by a valid link \lstinline{http://dev.mysql.com/doc/refman/5.5/en/keywords.html} in 7 posts.
However, there are still 949 posts that reference this broken link.
A more general case is that the linked site changes the location of the resource.
For example, 384 posts share the links that reference to \lstinline{ant.apache.org} website with the path \lstinline{manual/CoreTasks} in history.
All links (i.e., 63) that reference to \lstinline{ant.apache.org} website with the path \lstinline{manual/CoreTasks} are broken links (i.e., the server responded with 404 response code).
152 posts replace the path \lstinline{manual/CoreTasks} with \lstinline{manual/Tasks}.
This is because the website change the location of the documentation related to \textit{Tasks}.\footnote{https://github.com/apache/ant/commit/61b4c00b3852083b0f81-586d6f78adf0bc3c7f6f}
We suggest that the future researchers could repair the broken links on Stack Overflow based on the revisions of links.

\subsubsection{Actionable Suggestions for Stack Overflow}

\textbf{To help viewers avoid broken links when browsing Stack Overflow, Stack Overflow should detect the broken links and mark the questions with broken links.}
In Section \ref{findings_mechanism}, we observe that viewers vote more on the posts without broken links than the posts with broken links after the links become broken.
This shows the negative impact of the broken links.
We encourage Stack Overflow to take action on these broken links.
More specifically, Stack Overflow could learn from Wikipedia\footnote{https://en.wikipedia.org/wiki/User:Dispenser/Checklinks} where links can be reviewed, replaced with a working or archive link, tagged, or removed.
On Stack Overflow, moderators could use a tool, e.g., W3C checklink\footnote{http://validator.w3.org/checklink}, and Xenu's Link Sleuth\footnote{http://home.snafu.de/tilman/xenulink.html}, to automatically scan links to identify the broken links.
After that, Stack Overflow could include an \textit{broken link} mark for the questions with broken links.
By doing so, Stack Overflow users can be aware of the broken links when browsing the Stack Overflow websites even before clicking them.

\textbf{To maintain the crowdsourced knowledge on Stack Overflow, Stack Overflow should develop mechanisms to encourage users (especially posts owners) to pay more attention to the broken links and make efforts to maintain any broken links}.
In Section \ref{imp_res}, we observe that only 1.7\% of the posts with broken links are notified by the viewers in comments and only 5.8\% of the broken links that are posted on the Stack Overflow history are removed.
After notified by the viewers in comments, only 14.3\% of the posts ever with broken links repair the broken links.
This shows that the Stack Overflow gamification system (i.e., vote and accept answers) fails to encourage users to comment out and update broken links on Stack Overflow.
We suggest the Stack Overflow moderators could adjust the gamification system to encourage users to identify and update broken links.
For example, Stack Overflow could reward badges or reputation scores to users who identify or maintain broken links.

\textbf{Stack Overflow could archive snapshots of links as soon as they were created.}
In Section \ref{findings_websites}, we observe that the links that reference to the websites that host the resources that can be maintained by users are the least maintained.
These resources can be customized created, maintained, and deleted by their users.
More specifically, Stack Overflow could learn from Google\footnote{https://support.google.com/websearch/answer/1687222} and the Internet Archive's Wayback Machine\footnote{https://archive.org/web/} to archive snapshots of links when the links are posted.
Google takes a snapshot of each web page as a backup in case the current page is not available.
Internet Archive provides free universal access to books, movies, music, as well as 458 billion archived web pages.
However, these external web archive services cannot capture all the content of all the links at the time when the links are posted.
For example, a comment to the accepted answer that provides AI tools/frameworks/Library for Objective C\footnote{https://stackoverflow.com/q/5533317/} complains that:

\vspace{0.1cm} 
\hangindent 1.5em\textit{The A* link appears to be dead. The WayBack machine captured part of the post, although it missed most (only has DemoView.m and a small blurb remain). \underline{web.archive.org/web/20090207003416/http://bravobug.com/n}-\underline{ews/… }\footnote{http://web.archive.org/web/20090207003416/http://bravobug.c-om/news/?p=118}}
\vspace{0.1cm}
  
\noindent\normalsize
This comment shows that the external web archive service only captures part of the content\footnote{http://bravobug.com/news/?p=118}.
Stack Overflow could periodically replace the broken links on Stack Overflow with the links to the copies of the resources in the archive.

\subsubsection{Actionable Suggestions for Users}

\textbf{Stack Overflow users are encouraged not to remove the examples in the links in Stack Overflow questions.}
In Section \ref{findings_role}, we observe that 65\% of the broken links in our sampled questions are used to show examples, e.g., code examples.
This shows that the examples hosted in the links in the questions is removed after the problems are resolved.
However, this practice would lead to these questions with broken links to be totally useless as no following viewers can understand the questions.
We suggest Stack Overflow users not remove the examples of the links, especially in questions.

\textbf{We recommend that Stack Overflow users post the code in a more permanent site, e.g., code blocks and Stack Snippets on Stack Overflow as much as possible, rather than the ephemeral external code websites, e.g., github.com}.
In Section \ref{findings_websites}, we observe that code websites host the largest number of broken links.
This shows that the links that reference on code websites on Stack Overflow are often ephemeral (they get broken after some time has passed).
Stack Overflow provides code blocks for users to paste code snippets, and even Stack Snippets enable users to post runnable code \cite{code_editing, stack_snippets}.
We strongly recommend Stack Overflow users to post the code in the code blocks and Stack Snippets as much as possible, rather than external code websites.

\subsubsection{Feedback from Stack Overflow}
To understand whether our research can characterize broken links problems and obtain useful findings for Stack Overflow, we shared our findings with Stack Overflow community\footnote{https://meta.stackexchange.com/q/353998/}.
They concurred with our findings and see the importance of broken links problems on Stack Overflow (\textit{Six comments complain about the broken links. That is six missed opportunities to fix those links ...} – rene, \textit{I really admire your time and dedication, great effort.}  – Shadow Wizard Wearing Mask).
They saw the values of our findings and implications (\textit{``We suggest SO should directly archive the links when links are posted.'' In addition to keeping links alive, this also solves the problem of all that reviewer time wasted on deleting link-only answers. Win!} – francescalus).
They were interested in our work and requested for more details.
For example, they requested us to present our research related to the proportion of broken links among the links that were posted each month.
Based on our findings, the Stack Overflow community was also interested in investigating the soft 404 problems (\textit{Many links redirect to some generic page (may or may not be covered by response codes) or show content like "Content could not be found" or similar. In other words, many sites try to hide the fact that a link is broken. For instance, if the returned page is very short it could be counted as a broken link (though some may contain a link to a new location). Or some commonly used phrases could be detected.}).
We suggest future research efforts should continue working with the Stack Overflow team to solve/alleviate the broken links problem

\subsection{Threats to Validity}\label{threats}

\vspace{0.1cm}\noindent\textbf{Threats to internal validity} concern the factors that could have influenced our results.
We heavily depend on manual processes as described in Section \ref{findings_role} and Section \ref{findings_websites}.
Like any human activity, our manual labeling process is subject to personal bias.
To reduce personal bias in the manual labeling process, each website was labeled by two of the authors and discrepancies were discussed until a consensus was reached.
We also showed that the level of inter-rater agreement of the qualitative studies is high (i.e., the values of Cohen's kappa ranged are between 0.69 to 0.96).

In Section \ref{findings_mechanism}, to estimate the vote scores that are received before the links being broken, we consider the detected broken links were \textbf{not} broken in the first 30 days after the post of the links.
However, the detected broken links could become broken in the first 30 days after the post of the links.
To get the proportion of broken links that have been broken in the first 30 days, we calculate the proportion of broken links among the links that were posted from Aug. 26, 2020, to Sept. 26, 2020.
We obtain 89,632 posts and comments that were posted with external links from Aug. 26, 2020, to Sept. 26, 2020 from the Stack Exchange Data Explorer.
Then, we randomly sampled a statistically representative sample of 661 posts and comments with links from the 89,632 posts and comments, using a 99\% confidence level with a 5\% confidence interval.
We visit the sampled posts and comments and randomly click one of the links in the posts and comments and find 4 broken links, i.e., 0.6\% of the links that were posted in 20 days were broken links.
This indicates that only 6\% (i.e., 139,427) of broken links have been broken in the first 30 days.
This gives us high confidence in using the vote to the posts with broken links in the first 30 days after being posted to estimate the votes that are received before the link becomes broken.

In Section \ref{findings_mechanism}, to estimate the vote scores that are received after the links being broken, we consider the detected broken links were broken before the last 30 days before the collection of the dataset.
However, the broken links could not become broken before the last 30 days before the collection of the dataset, e.g., being broken one day before the collection of the dataset.
This could make our estimation of the vote scores that are received after the links being broken involve the votes that are received before the links being broken (i.e., a different type).
To identify the broken links that have not been broken in the last 30 days, we count the number the broken links that are only identified in the recent link availability test (i.e., Jan 2020).
As is indicated in Section 3.2, the time interval between two link availability tests is one month.
As a result, we observe 62,441 broken links that are newly observed in Jan 2020.
This indicates that 97.5\% (i.e., 2,430,887) of the broken links have been broken links in the last 30 days before the collection of the dataset.
This gives us high confidence in using the vote to the posts with broken links in the last 30 days before the collection of the dataset to estimate the votes that are received after the link becomes broken

\vspace{0.1cm}\noindent\textbf{Threats to external validity} concern the generalization of our findings.
Our study is conducted to investigate the broken links on Stack Overflow.
That said, our findings may not be generalized to the broken links in other Q\&A sites.
For example, other Q\&A forums that focus on a particular technology, e.g., Google Product Forums\footnote{https://productforums.google.com/forum/} and Microsoft Community\footnote{https://answers.microsoft.com/en-us/}, only share the links that relate to the specific technology.
In contrast, Stack Overflow is a popular website for developers and covers a wide range of programming-related technologies, and the links are prevalently shared across technologies.
In the future, we plan to analyze broken links in other Q\&A systems.

\section{Conclusion}\label{con}

In this paper, we investigate the broken links on Stack Overflow.
As a result, we observe that there are 1,687,995 broken links (14.2\%) in the latest version of Stack Overflow posts.
65\% of the broken links in our sampled questions are used to show examples, e.g., code examples.
70\% of the broken links in our sampled answers are used to provide supporting information, e.g., explaining a certain concept.
Only 1.57\% of the posts with broken links are highlighted as such by viewers in the posts' comments.
Only 5.8\% of the posts with broken links removed the broken links.
Viewers cannot fully rely on the vote scores to detect broken links, as broken links are common across posts with different vote scores.
The websites that host resources that can be maintained by their users are referenced by broken links the most on Stack Overflow, e.g., github.com.
Web technology related questions, e.g., JavaScript, HTML, CSS, and jQuery, are more likely to have broken links.

In the future, we plan to design a tool to repair the broken links on Stack Overflow.

\balance
\bibliographystyle{IEEEtran}
\bibliography{myreference}
\end{document}